\pdfoutput=1
\documentclass[11pt,twoside,a4paper,cmspaper,final,collab]{cms-tdr}
\def\svnVersion{d45f771}\def\svnDate{2024/10/10}\def\cmsCernNoTag{CERN-EP-2024-137}\def\cmsCernDate{\today}\def\cmsMessage{Published in Reports on Progress in Physics as \href{http://dx.doi.org/10.1088/1361-6633/ad7e4d}{\doi{10.1088/1361-6633/ad7e4d}.}}
\begin{document}\cmsNoteHeader{TOP-23-001}

\newlength\cmsTabSkip\setlength{\cmsTabSkip}{1ex}
\ifthenelse{\boolean{cms@external}}{%
    \providecommand{\cmsLeft}{upper\xspace}
    \providecommand{\cmsRight}{lower\xspace}
    \providecommand{\cmsTable}[1]{\resizebox{\linewidth}{!}{#1}}
}{%
    \providecommand{\cmsLeft}{left\xspace}
    \providecommand{\cmsRight}{right\xspace}
    \providecommand{\cmsTable}[1]{#1}
}

\newcommand{\PGht}{{\HepParticle{\PGh}{\PQt}{}}\xspace}
\newcommand{\Pellp}{{\HepParticle{\Pell}{}{+}}\xspace}
\newcommand{\Pellm}{{\HepParticle{\Pell}{}{-}}\xspace}
\newcommand{\pp}{\ensuremath{\Pp\Pp}\xspace}
\newcommand{\gluglu}{\ensuremath{\Pg\Pg}\xspace}
\newcommand{\ellell}{\ensuremath{\Pellp\Pellm}\xspace}
\newcommand{\ee}{\ensuremath{\Pe\Pe}\xspace}
\newcommand{\emu}{\ensuremath{\Pe\PGm}\xspace}
\newcommand{\mumu}{\ensuremath{\PGm\PGm}\xspace}
\newcommand{\Zjets}{\ensuremath{\PZ{+}\text{jets}}\xspace}
\newcommand{\tW}{\ensuremath{\PQt\PW}\xspace}
\newcommand{\ttV}{\ensuremath{\ttbar\PV}\xspace}

\newcommand{\mtop}{\ensuremath{m_{\PQt}}\xspace}
\newcommand{\mW}{\ensuremath{m_{\PW}}\xspace}
\newcommand{\mttbar}{\ensuremath{m(\ttbar)}\xspace}
\newcommand{\LQCD}{\ensuremath{\Lambda_{\text{QCD}}}\xspace}
\newcommand{\cosphi}{\ensuremath{\cos\varphi}\xspace}
\newcommand{\betaztt}{\ensuremath{\beta_z(\ttbar)}\xspace}
\newcommand{\Irel}{\ensuremath{I_{\text{rel}}}\xspace}
\newcommand{\pttort}{\ensuremath{\pt(\PQt/\PAQt)}\xspace}
\newcommand{\cosTheta}{\ensuremath{\cos\Theta}\xspace}
\newcommand{\mll}{\ensuremath{m(\Pellp\Pellm)}\xspace}

\newcommand{\trace}{\ensuremath{\mathrm{tr}}\xspace}
\newcommand{\hatlp}{\ensuremath{\hat{\Pell}{}^+}\xspace}
\newcommand{\hatlm}{\ensuremath{\hat{\Pell}{}^-}\xspace}
\newcommand{\hatlpm}{\ensuremath{\hat{\Pell}{}^{+/-}}\xspace}
\newcommand{\polvecp}{\ensuremath{\mathbf{B^+}}\xspace}
\newcommand{\polvecm}{\ensuremath{\mathbf{B^-}}\xspace}
\newcommand{\polvecpm}{\ensuremath{\mathbf{B^\pm}}\xspace}

\newcommand{\TTplusETAT}{\ensuremath{\ttbar{+}\PGht}\xspace}

\newcommand{\Pythia}{\ensuremath{\PYTHIA8}\xspace}
\newcommand{\Powhegvtwo}{\ensuremath{\POWHEG\textsc{v2}}\xspace}
\newcommand{\MGaMCatNLOOnly}{\ensuremath{\text{MG5\_a\MCATNLO}}\xspace}
\newcommand{\MGaMCatNLO}{\ensuremath{\MGaMCatNLOOnly\text{ (FxFx)}}\xspace}
\newcommand{\PowhegPythia}{\ensuremath{\Powhegvtwo{+}\Pythia}\xspace}
\newcommand{\MadgraphPythia}{\ensuremath{\MGaMCatNLO{+}\Pythia}\xspace}
\newcommand{\PowhegHerwig}{\ensuremath{\Powhegvtwo{+}\HERWIGpp}\xspace}
\newcommand{\PowhegPythiaEtatLong}{\ensuremath{\Powhegvtwo{+}\Pythia{+}\PGht}\xspace}
\newcommand{\PowhegPythiaEtat}{\ensuremath{\text{PH}{+}\text{P8}{+}\PGht}\xspace}
\newcommand{\PowhegHerwigEtat}{\ensuremath{\text{PH}{+}\text{Hpp}{+}\PGht}\xspace}

\cmsNoteHeader{TOP-23-001}
\title{Observation of quantum entanglement in top quark pair production in proton-proton collisions at \texorpdfstring{$\sqrt{s}=13\TeV$}{sqrt(s) = 13 TeV}}
\titlerunning{Quantum entanglement in top quark pair production}

\date{\today}

\abstract{
Entanglement is an intrinsic property of quantum mechanics and is predicted to be exhibited in the particles produced at the Large Hadron Collider. A measurement of the extent of entanglement in top quark-antiquark (\ttbar) events produced in proton-proton collisions at a center-of-mass energy of 13\TeV is performed with the data recorded by the CMS experiment at the CERN LHC in 2016, and corresponding to an integrated luminosity of 36.3\fbinv.
The events are selected based on the presence of two leptons with opposite charges and high transverse momentum. An entanglement-sensitive observable $D$ is derived from the top quark spin-dependent parts of the \ttbar production density matrix and measured in the region of the \ttbar production threshold. Values of $D<-1/3$ are evidence of entanglement and $D$ is observed (expected) to be $-0.480^{+0.026}_{-0.029}$ ($-0.467^{+0.026}_{-0.029}$) at the parton level. With an observed significance of 5.1 standard deviations with respect to the non-entangled hypothesis, this provides observation of quantum mechanical entanglement within \ttbar pairs in this phase space. This measurement provides a new probe of quantum mechanics at the highest energies ever produced.
}

\hypersetup{
pdfauthor={CMS Collaboration},
pdftitle={Observation of quantum entanglement in top quark pair production in proton-proton collisions at sqrt(s) = 13 TeV},
pdfsubject={CMS},
pdfkeywords={CMS, top quark, entanglement}}

\maketitle

\section{Introduction}

Entanglement is a fundamental concept in quantum mechanics that describes a strong correlation between particles, such that the state of one particle cannot be independently described without considering the state of the other, regardless of the distance between them~\cite{PhysRevLett.28.938, PhysRevLett.49.1804, PhysRevLett.89.240401}. Quantum entanglement has been extensively studied in the context of photons and electrons~\cite{Aspect1982, LoopholeFreeElectrons, LoopholeFreePhotons}. In these measurements, entanglement is often determined through observations of particle properties, such as spin or polarization~\cite{PfaffQbits}.
When considering the entanglement of two spin-1/2 particles, it is convenient to describe the quantum state of each particle as a linear combination of its basis states, usually denoted as $\vert0\rangle$ and $\vert1\rangle$ (corresponding to spin-up and spin-down states). In this formalism, one typically refers to each particle as a quantum bit (qubit) of information.
An entangled state for two qubits is a superposition of their joint states that cannot be factorized into individual states.
A common example is the Bell state, $\vert\psi^{-}\rangle=(\vert01\rangle-\vert10\rangle)/\sqrt{2}$, which can be realized to show the violation of Bell's inequality~\cite{PhysicsPhysiqueFizika.1.195, RoweViolationBell, AnsmannJoesephson}. This state provides important insights into the Einstein--Podolsky--Rosen paradox~\cite{PhysRev.47.777}, contributing to the ongoing discussion about the foundations of quantum mechanics. 
In a maximally entangled state, measuring the state of one qubit instantaneously determines the state of the other, independent of the physical separation between them. The theoretical exploration of entanglement
has recently been extended to collider settings involving a variety of fundamental particles including quarks, vector bosons, and the Higgs boson~\cite{Barr2023belltype, BARR2022136866, PhysRevD.107.076016, PhysRevD.107.016012, PhysRevD.107.093002, fictitiousTao, semileptonTao, bellChime, VARMA2024122795, Barr:2024djo, Fabbrichesi:2023cev, MaximizingEntBell}, as well as previously between taus produced from Z boson decays~\cite{ABEL1992304}.
Observation of quantum entanglement and violation of Bell's inequality has also been established in neutral B meson decays~\cite{PhysRevD.109.L031104}. 

Recently, the ATLAS Collaboration reported the first observation of entanglement in the top quark-antiquark (\ttbar) system~\cite{ATLAS:2023fsd}, with a result indicating a slight deviation from the standard model (SM) expectation.
We focus on a measurement of entanglement in top quark pairs, which provides a new probe to test predictions of quantum mechanics at the highest energies currently accessible~\cite{Afik:2020onf, Fabbrichesi:2021npl, SeveriQuantumTopsLHC}. This result differs from the result reported by the ATLAS Collaboration in that we measure the entanglement at the parton level while the ATLAS Collaboration reports their observable at the particle level. Furthermore, the result presented here considers non-relativistic bound-state effects in the production threshold for the first time. The formation of such bound states can be linked to the Sommerfeld enhancement~\cite{Roberto_Iengo_2009}, which describes the increase in annihilation cross section due to the non-relativistic attractive potential between particles. In the context of \ttbar production, the Sommerfeld enhancement arises from the resummation of ladder diagrams involving gluon exchanges, which is partially accounted for in next-to-leading order (NLO) corrections. However, NLO corrections also include diagrams besides ladder diagrams such as the emission of real particles. The CMS result rests on performing a binned likelihood fit to extract the entanglement proxy instead of utilizing a calibration curve.

The SM of particle physics is expected to break down at extremely short distances since gravity needs to be reconciled with quantum mechanics at the Planck scale. Such a short distance is beyond the reach of particle colliders, but this result sheds light on quantum correlations at distances corresponding to energies of hundreds of \GeVns. Furthermore, measurements involving intrinsic spin are a powerful probe of beyond the SM (BSM) contributions to the production of \ttbar pairs in proton-proton (\pp) collisions~\cite{MammenAbraham:2022yxp, Goncalves:2021dcu, PhysRevD.106.055007, PhysRevD.108.075027,Severi:2022qjy, FabbrichesiNewPhysicsInEntanglementTT, Maltoni:2024tul} and, possibly, at future circular lepton colliders~\cite{maltoni2024quantum}. Many of these BSM models alter the spin information of \ttbar pairs in complex and sizeable ways and, \eg, it has been shown in Ref.~\cite{PhysRevD.106.055007} that extensions of the SM implemented as an effective field theory can lower the level of entanglement of \ttbar pairs by up to 40\% relative to the SM expectation. In contrast, BSM effects can contribute up to a 15\% difference in the level of spin correlations of \ttbar pairs in the region $m(\ttbar) > 600$ \GeV, compared to the SM value in the same region. Hence, measurements of entanglement also provide a sensitive new probe of BSM contributions and represent a step towards quantum tomography and coherence measurements in top quark events~\cite{PhysRevD.106.055007}.

In this article, a measurement of the entanglement of \ttbar events produced in \pp collisions recorded with the CMS detector at a center-of-mass (CM) energy of 13\TeV at the CERN LHC is presented.
The top quark is the heaviest known fundamental particle and has a very short lifetime.
Expressed in natural units, the top quark's decay width $\Gamma_{\PQt}=1.42^{+0.19}_{-0.15}\GeV$~\cite{ParticleDataGroup}
is larger than the quantum chromodynamics (QCD) hadronization scale ($\LQCD\sim250\MeV$) and much larger than the spin decorrelation scale ($\LQCD^2/\mtop\sim0.36\MeV$, where \mtop is the mass of the top quark)~\cite{PhysRevD.81.074024}.
With such a large difference between the time scales of the top quark decay, the hadronization, and the spin decorrelation, the top quark spin retains its value from the moment it was created via the strong interaction.
Thus, top quarks provide unique direct access to their spin information, which is carried by the angular distributions of their decay products. These angular distributions provide a probe of the entanglement of top quarks and the measurement presented here utilizes the same data set and treatment of systematic uncertainties as in Ref.~\cite{CMS:2019nrx}, but with an optimized extraction technique to determine the extent of entanglement of top quarks and improved description of the production threshold region.

This article is structured as follows: Section~\ref{toc:threshold} covers the production and decay of top quarks, focusing on the \ttbar production threshold region. Section~\ref{toc:entanglement} explains the concept and significance of entanglement in top quark pairs, details the observable used to measure entanglement, and discusses the rationale for selecting the phase space region for this measurement. Section~\ref{toc:detector} outlines the CMS detector and provides details on identification and reconstruction of particles, while Sections~\ref{sec:dataset} and~\ref{sec:bkgestimate} provide information on the samples, analysis, and the background estimation. Section~\ref{sec:efficiencies} discusses the systematic uncertainties, and Section~\ref{sec:productionThreshold} introduces the pioneering modeling of the non-perturbative component of the \ttbar production threshold used in this analysis. Section~\ref{sec:extractcoefficients} explains the technique for statistically determining whether \ttbar pairs are entangled. Results of the measurement are discussed in Section~\ref{sec:results}, and a summary is given in Section~\ref{sec:summary}.

\section{The production and decay of top quarks}
\label{toc:threshold}

\begin{figure*}[!ht]
\centering
\includegraphics[width=0.75\textwidth]{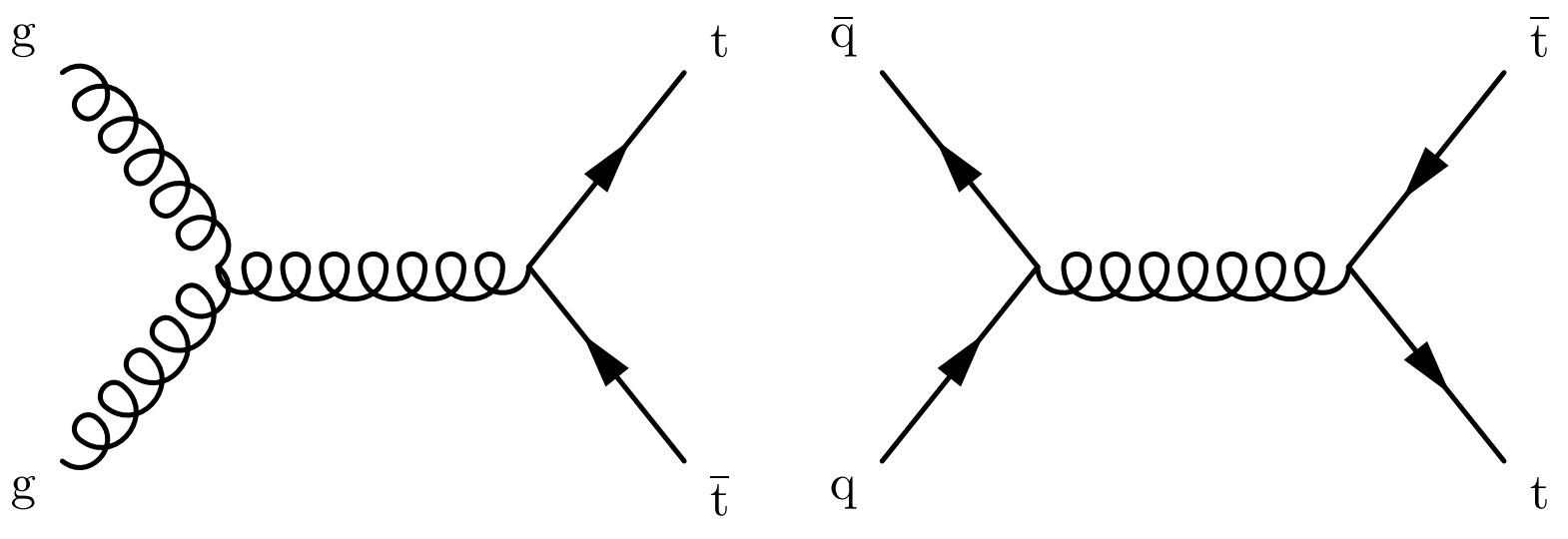}
\caption{%
    Representative leading order QCD Feynman diagrams for the \ttbar production through \gluglu fusion (left) and quark-antiquark annihilation (right).
}
\label{fig:feynman}
\end{figure*}

At the LHC, top quarks are produced predominantly in pairs (\ttbar) via the gluon-gluon (\gluglu) fusion process and at a level of about 10\% by quark-antiquark (\qqbar) annihilation, both mediated by the strong interaction, as described with perturbative QCD calculations~\cite{Czakon:2011xx, Czakon:2012zr, Czakon:2012pz, nnlo_theory}.
Figure~\ref{fig:feynman} shows the representative leading order (LO) Feynman diagrams for \ttbar production via the strong force.

Top quarks are mostly unpolarized when produced via the strong force, but their spins are strongly correlated. The configuration of the top quark-antiquark spin is dependent on the invariant mass of the \ttbar pair (\mttbar). The alignment (misalignment) of the spins of a top quark and a top antiquark results in like (unlike) helicity pairs, which dominate the low (high) invariant mass region. Top quark pairs originating from \gluglu fusion at the production threshold correspond to the Bell state, \ie, they are maximally entangled spin-singlets, while \qqbar annihilation produces \ttbar pairs that are separable at the production threshold~\cite{Afik:2020onf}. Thus, in the production threshold region \ttbar is produced in a mixed spin state. For a pure state, entanglement is a binary condition in which the pure state is non-separable. In contrast, for a mixed state, the degree of entanglement takes on a continuous value, measured via an entanglement witness or proxy, depending on the statistical ensemble of the individual states composing the mixed state. In this article, we exploit the \ttbar production threshold or ``low mass'' region to carry out a measurement of the entanglement of top quarks using an entanglement proxy.

Just below the \ttbar production threshold of $2\mtop\approx345\GeV$,
where sensitivity to \ttbar entanglement is enhanced, QCD allows the formation of a so-far-unobserved
bound state of a top quark and top antiquark, referred to as ``toponium''~\cite{PhysRevD.104.034023}.
Such states are a prediction of the SM following from the experimentally confirmed existence of charmonium and bottomonium bound states~\cite{ParticleDataGroup}.
Toponium bound states have yet to be discovered and they are not included in NLO QCD Monte Carlo (MC) simulations as toponium is inherently a non-perturbative phenomenon.
An observed toponium resonance would be the sum of the ground state, \PGht, and all other bound states, which are formed via nonrelativistic exchanges of gluons~\cite{NRQCD} that generate a Coulomb-like interaction between the top quarks.
Including toponium in \ttbar production simulations is important in light of the recent ATLAS and CMS differential cross section results~\cite{arXiv:2402.08486, CMS:2018adi, CMS:2016oae, ATLAS:2018orx, ATLAS:2020aln}, which indicate difficulties in modeling the data near the \ttbar production threshold.
The inclusion is also critical to this analysis as the pseudoscalar nature of toponium would lead to maximally entangled decay products.

Toponium production represents a distinct stage in the interaction dynamics between a top quark and antiquark, with both resonant color-singlet and nonresonant color-octet contributions. The total cross section---summed over all resonant and nonresonant contributions---is expected to peak at \mttbar = 400\GeV and to fall off until around 700\GeV~\cite{Kiyo:2008bv}. The color-singlet contribution is entirely dominated by the \gluglu initial state and is predicted to form a pseudoscalar resonance, the resonant \PGht state~\cite{Ju:2020otc}. In the \ttbar production threshold region, the \PGht state is predicted to form with a mass of 343\GeV
and a production cross section of $6.43\pm0.90\unit{pb}$~\cite{PhysRevD.104.034023,Sumino2010}.
The width of the predicted \PGht state, $\Gamma_{\PGht}=7\GeV$, is determined from the difference between the QCD predictions with and without Coulombic corrections~\cite{PhysRevD.104.034023}.
The \PGht color-singlet contribution is not negligible when compared to the \ttbar cross section near the production threshold~\cite{Maltoni:2024tul}. This is accounted for in this measurement by including a detailed study on the effect of the \PGht state on the measurement of the entanglement present in \ttbar events.

In addition to this color-singlet resonant contribution, there are also color-octet contributions to toponium production~\cite{Kiyo:2008bv}, which become more significant at longer distances. These are dominated by \gluglu initiated nonresonant color-octet pseudoscalar contributions and, to a smaller extent, \qqbar-initiated resonant color-octet vector contributions. However, the color-octet contribution is subject to a repulsive potential and only forms a bound state after an additional gluon is either absorbed or emitted~\cite{Fadin1990}. The latter complicates theoretical calculations of the octet contribution since these gluons are ``soft'' and translate to a large value of the strong coupling constant (\alpS).
Hence, there is no comprehensive theory for toponium production and thus no MC simulation of the full bound-state spectrum.
Owing to the lack of a model implementing these color-octet contributions, they are not included in the toponium model utilized here, which also applies to any interference terms. The phase space of the measurement presented here is selected to suppress \qqbar initiated processes, hence reducing the vector color-octet contributions to toponium production. Based on the state-of-the-art knowledge of toponium~\cite{Ju:2020otc, Beneke2022}, we estimate a systematic uncertainty of the toponium contribution in this measurement, which is explained in more detail in Section~\ref{sec:efficiencies}. Furthermore, the predicted \PGht state is a spin-singlet and therefore is expected to produce maximally entangled decay products in the phase space relevant for this analysis.

Top quarks decay almost exclusively via the electroweak interaction into a \PWp boson and a bottom quark. Similarly, top antiquarks decay to a \PWm boson and a bottom antiquark. This analysis targets events with two leptonic \PW boson decays, referred to as the ``dilepton'' (\ellell) decay channel, where \Pell refers to electrons or muons, including those originating from the decay of \PGt leptons. Entanglement effects are probed through a spin-correlation analysis of the lepton pair.

The squared matrix element (ME) for \ttbar production and decay (with appropriate color and spin summation understood)~\cite{PhysRevD.53.4886, Bernreuther2015, PhysRevD.81.074024} can be written as
\begin{equation}\label{eq:bernreuther}
    \abs{\mathcal{M}\big(\qqbar/\gluglu\to\ttbar\to(\Pellp\PGn\PQb)(\Pellm\PAGn\PAQb)\big)}^2\approx\trace[PR\overline{P}].
\end{equation}
In Eq.~\eqref{eq:bernreuther}, $\trace$ is the trace, $R$ is the production spin density matrix related to on-shell \ttbar production, and $P$ and $\overline{P}$ are the decay spin density matrices for the top quark and top antiquark, respectively~\cite{PhysRevD.53.4886}.
The narrow intrinsic width of the top quark allows the factorization of production and decay terms. According to Ref.~\cite{Afik:2020onf}, the most general form for the spin density operator $\rho$ is given by:
\begin{equation}
   \rho=\frac{1}{4}\Big(I_4+\sum_i\big(B_i^+\sigma^i\otimes I_2+B_i^-I_2\otimes\sigma^i\big)+\sum_{i,j}C_{ij}\sigma^i\otimes\sigma^j\Big),
\end{equation}
where $I_n$ is the $n{\times}n$ identity matrix, $\sigma^i$ refers to the Pauli matrices corresponding to each particle, and the indices $i$ and $j$ correspond to the three quantization axes ($i, j = 1, 2, 3$). The spin density operator $\rho$ is related to $R$ as $\rho=R/\trace(R)$.
In the case of two spin-1/2 particles, in particular, in \ttbar production in \pp collisions, \polvecpm are 3-vectors that characterize the degree of top quark/antiquark polarization along each of the axes, and $C$ is a $3{\times}3$ matrix that characterizes the correlation between the \PQt and \PAQt spins along each pair of axes. Given the on average net-zero polarization of \ttbar pairs near the \ttbar production threshold, the \polvecpm vectors are not expected to be sensitive to entanglement in \ttbar production.

Working in the \ttbar CM frame, we use the helicity axis $\hat{k}$ defined by the top quark direction and the direction $\hat{p}$ of the incoming parton to define the direction perpendicular to the scattering plane $\hat{n}=(\hat{p}\times\hat{k})/\sin\Theta$, where $\Theta$ is the top quark scattering angle. The direction in the scattering plane mutually perpendicular to $\hat{k}$ and the transverse axis $\hat{n}$ is given by $\hat{r}=(\hat{p}-\hat{k}\cosTheta)/\sin\Theta$. Although more details can be found in Ref.~\cite{CMS:2019nrx}, for this article the relevant quantities for understanding the measurement are the scattering angle defined above and the opening angle between the charged decay leptons in their parent top quark rest frames, $\varphi$, which is independent of the coordinate system defined here.

Equation~\eqref{eq:bernreuther} can be translated into spin correlation observables in the coordinate system defined above.
The cosine of the angle between the two charged decay leptons in their respective parent top quark rest frames, $\cosphi=\hatlp\cdot\hatlm$,
has a linear relationship with the relative differential cross section
\begin{equation}
    \frac{1}{\sigma}\dd{\sigma}{\cosphi}=\frac{1}{2}(1-D\cosphi),
\end{equation}
where the slope parameter $D$ is the entanglement proxy considered in this analysis, which is described in Section~\ref{toc:entanglement}, and the spin analyzing power of the leptons is assumed to be unity~\cite{BRANDENBURG2002235}.

\section{Entanglement observable for top quark pairs}
\label{toc:entanglement}

In general, measurements of top quark properties (specifically angular correlations) have been a highly sensitive probe as to whether the experimentally observed top quark is indeed the SM top quark~\cite{CDF:2010yag}.
Some top quark properties can be expressed as differential cross sections, for example distributions of spin correlation variables and hence the spin of top quarks is accessible via such distributions. In the realm of quantum measurements, it is possible to quantify varying degrees of entanglement within systems, influenced by factors such as phase space configurations and potential BSM contributions. This variability allows for a nuanced exploration of the hierarchy of quantum correlations within the spin density matrix, which can be conceptually organized as: Bell's inequality violation $\subseteq$ Entanglement  $\subseteq$ Spin correlations.
Compared to standard differential cross section measurements, estimating the systematic uncertainties relevant to conditions of entanglement requires---as it will be explained later---studies of different sources of systematic uncertainty and
this can have quite a different impact on the measurement.

A quantum state of two subsystems, denoted $\alpha$ for the \PQt quark and $\beta$ for the \PAQt antiquark, is separable when $\rho$ can be expressed as a convex sum of tensor products of states between the respective subsystems, denoted as $\rho_i^{\alpha}$ and $\rho_i^{\beta}$. A necessary and sufficient condition for identifying entanglement in such a two-particle system is the Peres--Horodecki criterion~\cite{bib:PhysRevLett.77.1413, bib:HORODECKI19961}, which allows the identification of entanglement in systems composed of two particles by examining the partial transpose of $\rho$, given by $\rho^{\mathrm{T}_2}=\sum_ip_i\rho_i^\alpha\otimes(\rho_i^\beta)^{\mathrm{T}}$, where $p_i$ are the probabilities of the individual density operators. If $\rho^{\mathrm{T}_2}$ has any negative eigenvalues, then the state $\rho$ is entangled.
This criterion can be used to show that top quarks are entangled by measuring a proxy that takes on particular values when this criterion is met.

We consider the momenta of the two leptons in the CM frame of the \ttbar system and an additional boost into the frame of their respective parent top quark or antiquark. The momentum directions of the two leptons in these CM frames are labeled as \hatlp and \hatlm.
The four-fold angular distribution for the two charged leptons \hatlpm in the \PQt and \PAQt reference frames
is given by~\cite{Baumgart:2012ay}
\begin{equation}\label{eq:fulldist}
    \frac{1}{\sigma}\frac{\rd^4\sigma}{\rd\Omega_+\rd\Omega_-}=\frac{1}{(4\pi)^2}\Big(1+\polvecp\cdot\hatlp+\polvecm\cdot\hatlm+\hatlp\cdot\big(C\hatlm)\Big).
\end{equation}
The charged lepton direction is used as a proxy for the top quark spin and the lepton is suitable for this purpose, since it has a spin analyzing power of $\approx$1~\cite{BRANDENBURG2002235}, which means that there is a strong correlation
between the parent quark spin and the angular distribution of the lepton. For simplicity, the spin analyzing power factors are omitted in Eq.~\eqref{eq:fulldist}.

Both the polarization vectors \polvecpm and the spin correlation matrix $C$ depend on the specifics of the \ttbar kinematic variables.
However, in this measurement we do not utilize the polarization vectors \polvecpm, since there is no net polarization of top quarks at LO and thus they are not expected to be sensitive to the entanglement of \ttbar pairs. Instead, we focus on the diagonal elements of the $C$ matrix.
At the production threshold, \ttbar production via \gluglu fusion results in spin-singlet states,
which means the quark spins are maximally correlated along any axis and $C$ is diagonal: $C=\mathrm{Diag}(-1,-1,-1)$.
Above threshold, the relative contribution of \gluglu fusion resulting in spin-singlets decreases with increasing \mttbar, resulting in more mixed states and diagonal elements of $C$ that are less than one~\cite{HORODECKI1995340}. 

The Peres--Horodecki criterion relates an entanglement witness $\Delta\equiv -C_{33}+\abs{C_{11}+C_{22}}-1$ to a condition for entanglement~\cite{Afik:2020onf} of $\Delta>0$, where the third axis is taken as the spin quantization axis. The specific choice of which basis vector corresponds to the spin quantization axis can be chosen at will and ultimately results in four unique observables that can be used for entanglement measurements~\cite{Maltoni:2024tul, Aguilar-Saavedra:2022uye}.

The entanglement proxy, or $D$ coefficient, is related to the diagonal $C$ coefficients as:
\begin{equation}
    D=\trace[C]/3=(C_{11}+C_{22}+C_{33})/3,
\end{equation}
which yields the condition for observing entanglement:
\begin{equation}\label{eq:entanglementConditionD}
    \Delta+1=\trace[C]=-3D>1.
\end{equation}
The coefficient $D$ itself is a measure of the fraction of events where the spins of the \PQt quark and \PAQt antiquark are aligned. To demonstrate entanglement in top quark events, we compare the measured entanglement proxy $D$ with the $-1/3$ boundary for entanglement in Eq.~\eqref{eq:entanglementConditionD}~\cite{Afik:2020onf}.
Measuring $D<-1/3$ implies that the partial transpose of the density operator is not positive definite and therefore is nonseparable, which is following the criteria discussed in Ref.~\cite{bib:HORODECKI19961}.

\begin{figure*}[!ht]
\centering
\includegraphics[width=0.48\textwidth]{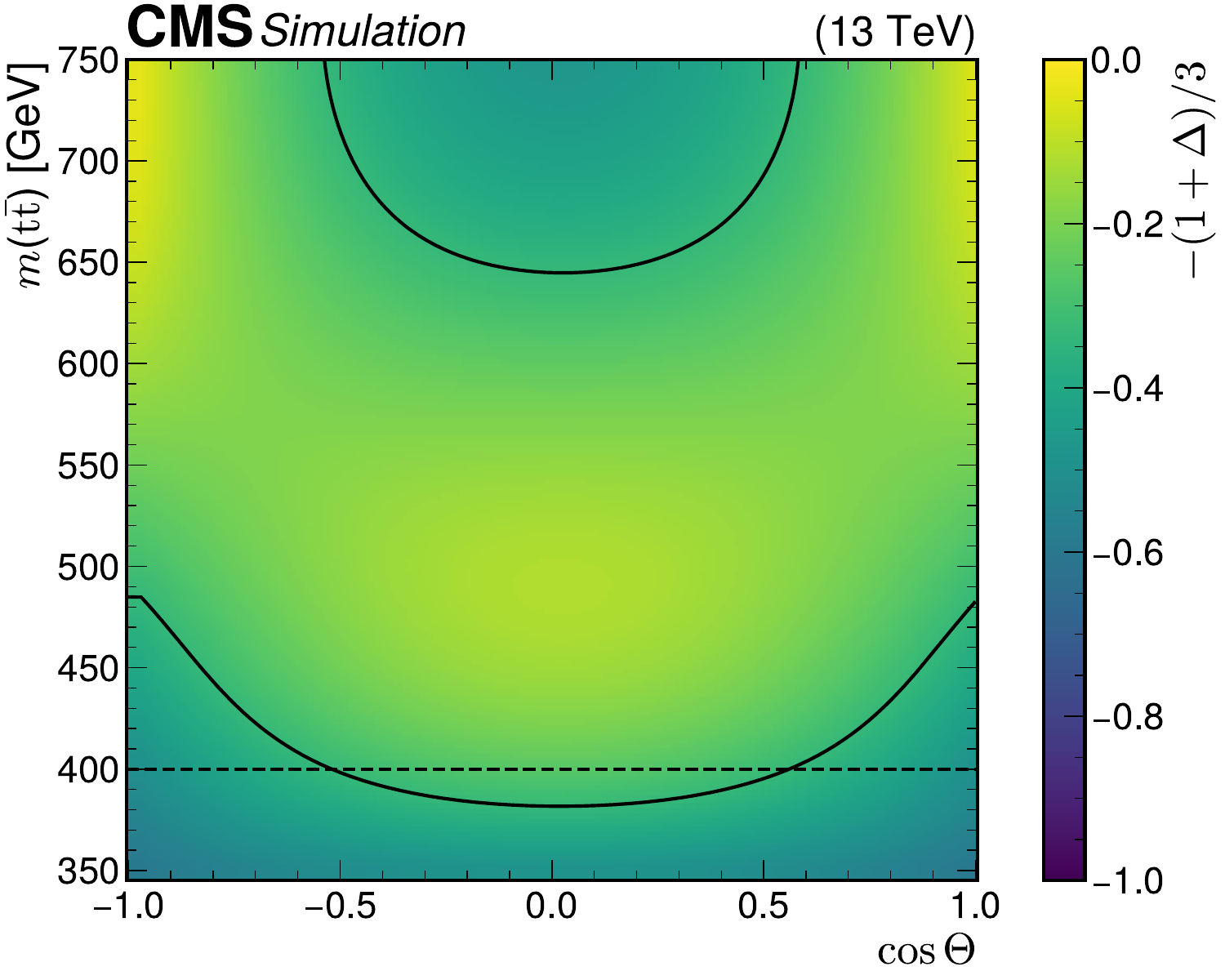}%
\hfill%
\includegraphics[width=0.48\textwidth]{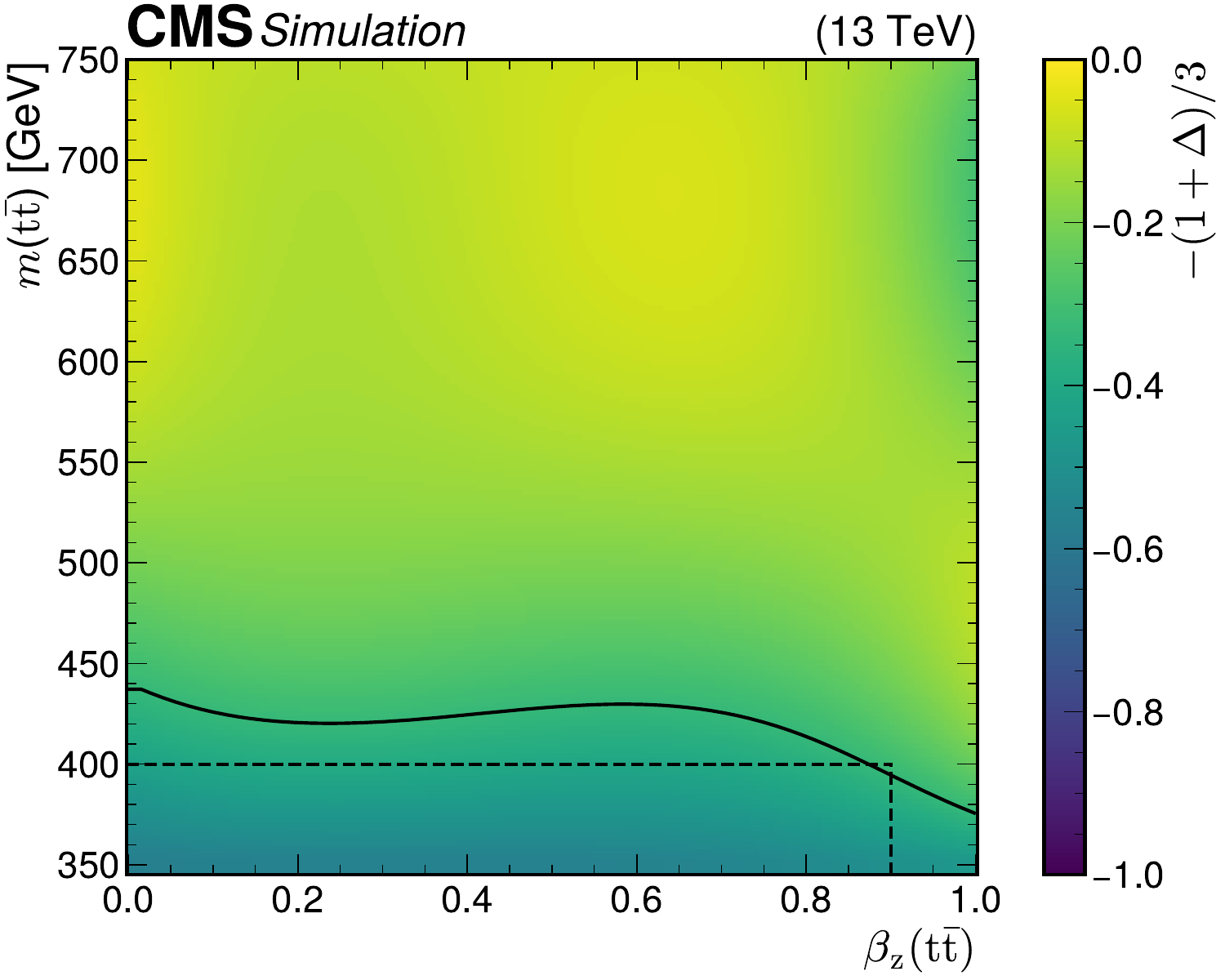}
\caption{%
    Predicted values of $-(1+\Delta)/3$ obtained from \ttbar MC simulation, without accounting for detector effects, are shown on the left as a function of \mttbar and the cosine of the top quark scattering angle $\Theta$.
    The value of $-(1+\Delta)/3$ also determined by a \ttbar MC simulation as a function of \mttbar and \betaztt is shown on the right.
    In both figures the black solid lines represent the boundary for entanglement, while the black dashed line indicates the selected phase space in this analysis. The minimum value on the color axis of $-1$ corresponds to the boundary $\trace[C]=3$, a maximally entangled state. Top quarks with no spin correlations correspond to a value of $D=0$ and $\Delta=-1$ ($C=\mathbf{0}$). In the central boosted region, $D<-1/3$ is not a valid entanglement criterion as the negativity of the $C_{11}$ and $C_{22}$ elements is not guaranteed~\cite{Aguilar-Saavedra:2022uye} and the correct entanglement criterion is $\Delta>0$.
}
\label{fig:money_plot}
\end{figure*}

Figure~\ref{fig:money_plot} (left) illustrates how the entanglement is expected to vary as a function of the kinematic parameters. The quantity $-(1+\Delta)/3$ is shown (on the color scale) as a function of \mttbar and the top quark scattering angle $\Theta$ determined using a \ttbar MC simulation implementing both the \gluglu fusion and \qqbar process at NLO. Intuitively, this means that at threshold, top quarks are produced in like-helicity states forming a spin-singlet state. This spin-singlet is maximally entangled and explains the large value of entanglement near the production threshold. Toward intermediate values of \mttbar, the level of entanglement is substantially reduced, while at very high \mttbar the unlike-helicity states dominate and form a spin triplet state, which is also maximally entangled.
We focus on the low \mttbar region, with $345<\mttbar<400\GeV$, where we expect both $C_{11}$ and $C_{22}$ to be negative~\cite{Bernreuther2015}.
This simplifies the $\Delta>0$ entanglement criterion to $\trace[C]>1$.

It was suggested in Ref.~\cite{Aguilar-Saavedra:2022uye} to enhance the purity of the \gluglu fusion process by an additional requirement on the relative longitudinal velocity between the lab and \ttbar reference frames: $\betaztt=\abs{(p_z^{\PQt}+p_z^{\PAQt})/(E^{\PQt}+E^{\PAQt})}$. Figure~\ref{fig:money_plot} (right) shows again the quantity $-(1+\Delta)/3$ (on the color axis) as a function of \mttbar and \betaztt. Below the dashed line indicates an entanglement sensitive phase space of $\mttbar<400\GeV$ and $\betaztt<0.9$.

For the measurement of entanglement in top quark events, we focus on a variable that can be measured more precisely~\cite{CMS:2019nrx}, \ie, $\cosphi=\hatlp\cdot\hatlm$, which fully encapsulates the differential spin correlation information at the production threshold through the \gluglu fusion process and in absence of BSM contributions~\cite{PhysRevD.106.055007}.

At the \ttbar production threshold, the \cosphi distribution provides the best sensitivity of all spin correlation variables and hence, the corresponding $D$ coefficient is least affected by systematic uncertainties~\cite{CMS:2019nrx, CMS-PAS-FTR-18-034}. The analysis strategy for measuring the entanglement in \ttbar pairs is based on using templates as a function of \cosphi to perform a binned profile likelihood fit with uncertainties as nuisance parameters using the CMS statistical analysis tool \textsc{combine}~\cite{CAT-23-001}, which is based on the \textsc{RooFit}~\cite{Verkerke:2003ir} and \textsc{RooStats}~\cite{Moneta:2010pm} frameworks. The final measured result in data is corrected to the parton level and compared to the boundary for entangled top quark-antiquark states, which is well defined as $D = -1/3$. Therefore, observation of top quark entanglement at the parton level is equivalent to a measurement of $D$ to be less than $-1/3$ within a given precision.

\section{The CMS detector and event reconstruction}
\label{toc:detector}

The central feature of the CMS apparatus is a superconducting solenoid of 6\unit{m} internal diameter, providing a magnetic field of 3.8\unit{T}. Within the solenoid volume are a silicon pixel and strip tracker, a lead tungstate crystal electromagnetic calorimeter (ECAL), and a brass and scintillator hadron calorimeter (HCAL), each composed of a barrel and two endcap sections. Forward calorimeters extend the pseudorapidity ($\eta$) coverage provided by the barrel and endcap detectors.
Muons are measured in gas-ionization detectors embedded in the steel flux-return yoke outside the solenoid, with detection planes made using three technologies: drift tubes, cathode strip chambers, and resistive plate chambers.
Events of interest are selected using a two-tiered trigger system~\cite{CMS:2016ngn}.
The first level, composed of custom hardware processors, uses information from the calorimeters and muon detectors to select events at a rate of around 100\unit{kHz} within a time interval of less than 4\mus.
The second level, known as the high-level trigger (HLT), consists of a farm of processors running a version of the full event reconstruction software optimized for fast processing, and reduces the event rate to around 1\unit{kHz} before data storage.
A more detailed description of the CMS detector, together with a definition of the coordinate system used and the relevant kinematic variables, can be found in Refs.~\cite{CMS:2008xjf,CMS:2023gfb}.

A particle-flow (PF) algorithm~\cite{CMS:2017yfk} aims to reconstruct and identify each individual particle in an event, with an optimized combination of information from the various elements of the CMS detector.
The primary vertex is taken to be the vertex corresponding to the hardest scattering in the event.
The energy of photons is obtained from the ECAL measurement. The energy of electrons is determined from a combination of the electron momentum at the primary interaction vertex as determined by the tracker, the energy of the corresponding ECAL cluster, and the energy sum of all bremsstrahlung photons spatially compatible with originating from the electron track. The energy of muons is obtained from the curvature of the corresponding track. The energy of charged hadrons is determined from a combination of their momentum measured in the tracker and the matching ECAL and HCAL energy deposits, corrected for the response function of the calorimeters to hadronic showers. Finally, the energy of neutral hadrons is obtained from the corresponding corrected ECAL and HCAL energies.

Jets are reconstructed by clustering the PF candidates using the infrared- and collinear-safe anti-\kt clustering algorithm with a distance parameter of 0.4~\cite{Cacciari:2008gp, Cacciari:2011ma}.
Jet momentum is determined as the vectorial sum of all particle momenta in the jet, and is found from simulation to be, on average, within 5--10\% of the true momentum over the whole transverse momentum (\pt) spectrum and detector acceptance. Additional \pp interactions within the same or nearby bunch crossings (pileup) can contribute additional tracks and calorimetric energy depositions, increasing the apparent jet momentum. To mitigate this effect, tracks identified to be originating from pileup vertices are discarded and an offset correction is applied to correct for remaining contributions~\cite{CMS:2020ebo}. Jet energy corrections are derived from simulation studies so that the average measured energy of jets becomes identical to that of particle level jets. In situ measurements of the momentum balance in dijet, photon$+$jet, {\PZ}$+$jet, and multijet events are used to determine any residual differences between the jet energy scale in data and in simulation, and appropriate corrections are made~\cite{CMS:2016lmd}.
Additional selection criteria are applied to each jet to remove jets potentially dominated by instrumental effects or reconstruction failures~\cite{CMS:2020ebo}. The jet energy resolution amounts typically to 15--20\% at 30\GeV and 10\% at 100\GeV~\cite{CMS:2016lmd}.

The missing transverse momentum vector \ptvecmiss is computed as the negative vector \pt sum of all the PF candidates in an event, and its magnitude is denoted as \ptmiss
\cite{CMS:2019ctu}. The \ptvecmiss is modified to account for corrections to the energy scale of the reconstructed jets in the event.

Electron candidates, reconstructed from a combination of the track momentum vector at the primary interaction vertex and the corresponding clusters in the ECAL, are excluded if the ECAL clusters are in the region between the barrel and endcap ($1.44<\abs{\eta_{\text{cluster}}}<1.57$), since they have a reduced reconstruction efficiency there.
A relative isolation criterion $\Irel<0.0588$ (0.0571) is applied for electron candidates in the barrel (endcap). The \Irel is defined as the \pt sum of all photon and neutral and charged hadron candidates within a distance of 0.3 from the electron candidate in $\eta$-$\varphi$ space, divided by the \pt of the electron candidate, with a correction to suppress the residual effect of pileup.
Additional electron identification requirements are applied to reject misidentified electron candidates and candidates originating from photon conversions~\cite{CMS:2017yfk, CMS:2015xaf}.
The electron momentum is estimated by combining the energy measurement in the ECAL with the momentum measurement in the tracker. The momentum resolution for electrons with $\pt\approx45\GeV$ from $\PZ\to\Pe\Pe$ decays ranges from 1.6--5\%. It is generally better in the barrel region than in the endcaps, and also depends on the bremsstrahlung energy emitted by the electron as it traverses the material in front of the ECAL~\cite{CMS:2020uim,CMS-DP-2020-021}.

Muon candidates are reconstructed using the track information from the tracker and the muon system~\cite{CMS:2018rym}.
A relative isolation requirement of $\Irel<0.15$ within a distance of 0.4 in $\eta$-$\varphi$ space from the muon candidate is applied.
To reject misidentified muon candidates and candidates originating from decay-in-flight processes, additional muon identification requirements are used~\cite{CMS:2018rym}.
The efficiency to reconstruct and identify muons is greater than 96\%. Matching muons to tracks measured in the silicon tracker results in a relative \pt resolution of 1 (3)\% in the barrel (endcaps) for muons with \pt up to 100\GeV.

\section{Data and simulated samples}
\label{sec:dataset}

The analysis is performed using 36.3\fbinv of \pp collision data collected by the CMS experiment during the 2016 LHC run at a CM energy of 13\TeV. The single muon trigger efficiency exceeds 90\% over the full $\eta$ range. To maximize the trigger efficiency, both single-lepton and dilepton triggers are used. Data events are required to pass one of the following dilepton or single lepton triggers. For events containing two reconstructed electrons with opposite charge, the \pt of the leading (sub-leading) electron is required to be larger than 23 (12)\GeV. For events containing two reconstructed oppositely charged muons, the \pt of the leading (sub-leading) muon is required to be larger than 17 (8)\GeV. For events containing a reconstructed electron and a muon, either the muon \pt is required to be greater than 23\GeV and the electron \pt greater than 12\GeV or the muon \pt is required to be greater than 8\GeV and the electron \pt greater than 23\GeV. The dilepton triggers additionally require the reconstructed $z$ coordinate of the vertex position of the leptons to be within 0.2\cm of each other. The single lepton triggers require the electron \pt to be greater than 27\GeV and the muon \pt greater than 24\GeV.

We utilize a combined simulated signal sample consisting of a (dominant) \ttbar contribution and the \PGht contribution, for the top quark bound state described in Section~\ref{toc:threshold}.
The \ttbar signal sample is produced using the \Powhegvtwo event generator at NLO~\cite{Frixione:2007nw, Nason:2004rx, Frixione:2007vw, Alioli:2010xd} including spin correlations and thus referred to as the ``spin correlated'' (SC) or ``nominal \ttbar'' sample.
The nominal \ttbar sample is reweighted to a \Powhegvtwo sample with spin correlations turned off differentially in $\mttbar\otimes\cosphi\otimes \betaztt\otimes\cosTheta$, with otherwise the exact same settings, to implement the \ttbar ``no spin correlation'' (noSC) case.
We rely on the \textsc{top++} v2.0 program~\cite{Czakon:2011xx} to calculate the \ttbar production cross section at next-to-NLO (NNLO) including electroweak (EWK) corrections at NLO, and obtain $832^{+20}_{-29}\,(\text{scale})\pm 35\,(\text{PDF}{+}\alpS)\unit{pb}$ assuming $\mtop=172.5\GeV$. In order to assess the level of variation when using an alternative ME and matching procedure, a second \ttbar sample is obtained using \MGvATNLO version 2.4.2 (\MGaMCatNLOOnly)~\cite{Alwall:2014hca} including \textsc{MadSpin}~\cite{Artoisenet:2012st} at NLO, where the ME jets are matched to parton showers using the FxFx prescription~\cite{Frederix:2012ps}. The parton distribution functions (PDFs) in all cases are described using NNPDF3.0~\cite{NNPDF:2014otw}, and \mtop is assumed to be 172.5\GeV.

The \PGht signal sample is produced using the \MGvATNLO version 2.6.5 event generator at LO~\cite{PhysRevD.104.034023} including spin correlations. The \PGht particle couples exclusively to gluons and top quarks and is produced through an $s$-channel process resulting in a \ttbar pair.
In the production threshold region, this calculation predicts the \PGht to form with a mass of 343\GeV and width of 7\GeV with a production cross section of 6.43\unit{pb}~\cite{PhysRevD.104.034023}. This \PGht signal sample is produced at parton level selecting only events with masses in between 337--349\GeV. The generator-level events in the simulated \PGht sample are smeared to the reconstruction level utilizing the nominal \ttbar sample.
To get a noSC \PGht sample, the nominal \MGvATNLO version 2.6.5 \PGht sample is reweighted using the same procedure that was applied to the nominal \Powhegvtwo~\ttbar sample.

The combined signal model of \TTplusETAT includes corrections to account for EWK contributions at NLO, referred to as ``EWK corrections", and the corrections to the average of the transverse momentum of the top quark and top antiquark (\pttort) spectra derived from NNLO QCD. For the \pttort reweighting, also referred to as ``NNLO QCD reweighting'', we use \textsc{fastNLO} tables~\cite{Czakon:2017dip, Czakon:2015owf, Czakon:2016dgf} to compute the \pttort distribution in fixed-order QCD at NNLO using the same PDF and \mtop value as was used in the \ttbar sample but excluding EWK contributions at NLO, since those are applied in a separate additional step. The EWK corrections are taken into account by following a similar approach as in Ref.~\cite{CMS:2020djy} to determine the size and uncertainty of these corrections (discussed in Section~\ref{sec:efficiencies}). The EWK corrections are computed in \textsc{hathor} at NLO differentially in \cosTheta and \mttbar.
The generator level values of \mttbar and \cosTheta are used to determine the appropriate $\kappa_{\text{NLO}}^{\text{EWK}}$ scale factor (SF) per event. Applying these EWK corrections on a bin-by-bin basis can be performed via two different methods: ``multiplicative'' or ``additive''. The multiplicative approach determines a SF, $\kappa_{\text{NLO}}^{\text{EWK}}$, that is applied multiplicatively to the nominal \ttbar sample, while the additive approach adds the predicted EWK contribution to the nominal \ttbar sample. In most regions of the kinematic phase space, the EWK contributions to the differential cross section factorize, and therefore the multiplicative approach is a good approximation to correct for higher-order EWK contributions. However, the difference between the two approaches is taken as a systematic uncertainty.

The major sources of background contributions are \Zjets, single top quark \tW, dibosons ($\PW\PW$, $\PW\PZ$, and $\PZ\PZ$), and \ttbar events in association with {\PW}/\PZ bosons (\ttV).
The \Zjets sample is simulated at NLO with \MGaMCatNLOOnly (FxFx matching)~\cite{Frederix:2012ps}, single top quark samples are produced at NLO with \Powhegvtwo, diboson events are simulated at LO with \PYTHIA8.219~\cite{Sjostrand:2014zea}, (referred to as \Pythia in what follows), and \ttV samples are produced at NLO with \MGaMCatNLOOnly (FxFx matching).

For all signal and background samples, the parton showering and hadronization is performed with \Pythia. While the \Pythia CUETP8M2T4 tune~\cite{CMS-PAS-TOP-16-021} is employed to describe the underlying event in the \ttbar, \PGht, and single top quark samples, all other background samples utilize the \Pythia CUETP8M1 tune~\cite{Khachatryan:2015pea}.
An alternative showering and hadronization model, \HERWIGpp~\cite{Bahr:2008pv}, is used to quantify the sensitivity to the modeling of these processes.
Where \HERWIGpp is used for parton showering and hadronization, the EE5C tune~\cite{Gieseke:2012ft} is employed instead.

Pileup collisions are overlaid to each simulated event, and the generated distribution of the number of events per bunch crossing is matched to that observed in data. The full CMS detector simulation is carried out using \GEANTfour version 9.4~\cite{AGOSTINELLI2003250}.

\section{Event selection, background estimation, and kinematic reconstruction}
\label{sec:bkgestimate}

The measurement of the entanglement proxy $D$ employs, to a large extent, the event selection and approaches to background estimation and corrections as in Refs.~\cite{CMS:2019nrx, CMS:2018adi}, which measured $D$ at the inclusive level to be $0.237\pm0.011$~\cite{CMS:2019nrx}, indicating no entanglement over the full phase space.
For this measurement, the \ttbar pairs are restricted to a phase space where the level of entanglement is expected to be high and the selection is optimized to enhance entangled top quarks as discussed later in this section.

Both electron and muon candidates are required to have $\pt>25$ (20)\GeV for the leading (trailing) candidate and $\abs{\eta}<2.4$. Electrons and muons originating from tau lepton decays are considered as signal for this analysis. Jets are selected if they have $\pt>30\GeV$ and $\abs{\eta}<2.4$ with additional isolation requirements.

Jets originating from the hadronization of \PQb quarks (\PQb jets) are identified (\PQb tagged) by combining information related to secondary decay vertices reconstructed within the jets and track-based lifetime information in an algorithm (CSVv2) that provides the \PQb jet identification efficiency of 79--87\% and a probability to misidentify light- and charm-flavor jets as \PQb jets of approximately 10 and 40\%, respectively~\cite{CMS:2017wtu}.

The selected events are required to have exactly two isolated electrons or muons of opposite electric charge and at least two jets. At least one of the jets is required to be \PQb tagged.
Events with a lepton-pair invariant mass $\mll<20\GeV$ are removed in order to suppress contributions from heavy-flavor resonance decays and low-mass Drell--Yan processes.
In the \EE and \MM channels, backgrounds from \Zjets processes are further suppressed by requiring $\ptmiss>40\GeV$ and vetoing events in the \PZ boson mass region, with $76<\mll<106\GeV$.
The remaining background yield from \Zjets events, which is large in the \EE and \MM channels, is determined by applying a factor derived from simulation to the number of \Zjets events observed in data in a control region where \mll is close to the \PZ boson mass~\cite{TOP-11-005_paper}.
A correction to account for non-\Zjets backgrounds in the control region is derived from the $\Pepm\PGmmp$ channel. The simulated \Zjets yield is corrected by a SF of up to 1.05 in each channel to match the determination from data.

Other background sources include processes of diboson, single top quark \tW, and $\ttbar{+}\PZ/\PW$ (\ttV) production, as well as misidentified \ttbar events originating from lepton$+$jets and hadronic decay channel events. These \ttbar events are collectively referred to as ``\ttbar other''. The same category for the \PGht only consists of 11 events. Hence, it is negligible compared to the \PGht signal component of 892 events and not included.

The number of {\PW}$+$jets events is found to be negligible ($<$0.5\%) and hence is not included as a background source. Overall, the event selection as described here has a purity of $>$88\%.

To determine the most sensitive phase space for measuring the entanglement proxy $D$ while minimizing expected total uncertainties, we scan the phase space of $\cosphi\otimes\mttbar\otimes\betaztt$. The region with the highest expected sensitivity is determined based on maximizing the expected sensitivity to observe entangled \ttbar pairs. We converge on a best choice of $345<\mttbar<400\GeV$ along with an additional requirement of $\betaztt<0.9$, shown in Fig.~\ref{fig:money_plot} with the dashed lines.
The final event yields displayed in Table~\ref{tab:yields1} are obtained after applying all selection requirements.

\begin{table}[!ht]
\centering
\topcaption{%
    The expected number of events from signal and background contributions after event selection, compared with the number observed in data. The ``\ttbar other'' category includes mis-identified semileptonic and fully hadronic decays, and hadronic decays of tau leptons of the \ttbar pairs. The uncertainties include only the MC statistical uncertainties. The ``Only \PGht'' contribution is not added to the total MC prediction since it is included in the combined (\TTplusETAT) signal contribution.
}
\label{tab:yields1}
\renewcommand{\arraystretch}{1.2}
\cmsTable{\begin{tabular}{c D{,}{\,\pm\,}{-1} D{,}{\,\pm\,}{-1}}
    \multirow{2}{*}{Sample}
    & \multicolumn{1}{c}{\multirow{2}{*}{Full phase space}}
    & \multicolumn{1}{c}{$345<\mttbar<400\GeV$}\\[-2pt]
    & & \multicolumn{1}{c}{\& $\betaztt<0.9$} \\
    \hline
    Signal (\TTplusETAT) & 230580,210 & 45793,92 \\
    \Zjets & 11300,400 & 3560,260 \\
    \tW & 8990,110 & 1873,49 \\
    \ttbar other & 1916,19 & 572,10 \\
    Diboson & 711,19 & 179.6,9.4 \\
    \ttV & 691.6,5.8 & 100.1,2.1 \\
    Total MC & 254190,500 & 52080,300 \\
    Only \PGht & 1554.4,6.9 & 892.2,6.0 \\[\cmsTabSkip]
    Data & 255646 & 53843 \\
\end{tabular}}
\end{table}

All of the background contributions except those from \Zjets are estimated from simulation following Ref.~\cite{CMS:2019nrx}.

The yields from simulation are normalized to the integrated luminosity of 36.3\fbinv.
Data-to-simulation SFs are applied to simulated events to account for trigger, lepton, \PQb tagging, and kinematic reconstruction efficiencies.

The four-momenta of the \PQt and \PAQt in each event are estimated using a kinematic reconstruction algorithm~\cite{CMS:2018adi, CMS:2015rld}.
The algorithm solves for the unknown neutrino momenta under these constraints: \ptmiss is attributed solely to the two neutrinos; the invariant mass of each \PW boson, \mW, is fixed to 80.4\GeV~\cite{ParticleDataGroup};
and the invariant mass of each top quark must equal 172.5\GeV. Detector resolution effects are modeled by smearing the measured energies and directions of jets and leptons based on their simulated resolutions, and \mW is smeared according to a relativistic Breit--Wigner distribution with a width of 2.1\GeV~\cite{ParticleDataGroup}.
The best-fit solution for neutrino momenta is selected based on the smallest reconstructed \mttbar, and solutions are weighted by the expected invariant mass spectrum of the lepton and \PQb jet system ($m(\Pell\PQb)$) from the top quark decay~\cite{CMS:2018adi} after the kinematic selection requirements.
Weights from 100 smearings are used to compute the four-momenta as a weighted average. This reconstruction is performed for all suitable pairs of \PQb jet candidates for the two possible lepton-jet assignments. Only if no kinematic solution is found at all, then also combinations with one \PQb-tagged and one untagged jet are considered. The jet pair and lepton-jet assignment that yields the maximum sum of weights is chosen. The kinematic reconstruction efficiency is about 90\% in both data and simulation; the remaining 10\% are events without a real neutrino momenta solution.

\section{Systematic uncertainties}
\label{sec:efficiencies}

In this section, each of the uncertainties in the measurement of the top quark entanglement proxy $D$ is discussed.
For each source of systematic uncertainty, templates as a function of \cosphi are generated based on $\pm1$ standard deviation ($\sigma$) uncertainty variations and Table~\ref{tab:syst_overview} shows a summary of systematic uncertainties and their impact on the measurement at the pre-fit level. The templates are used as inputs for a binned profile likelihood fit.

\begin{table*}[!tp]
\centering
\topcaption{%
    An overview of the systematic uncertainties and their impact on the yields and shape of the \cosphi distribution. The uncertainties are categorized by their type where ``norm.'' refers to normalization uncertainties modeled with a log-normal prior and ``shape'' refers to shape uncertainties. The impact on the yields and shape of the \cosphi distribution is given in percent where the difference in the shape of the \cosphi distribution is determined from the forward-backward asymmetry. The JES systematics are split as in Ref.~\cite{CMS:2016lmd} with the addition of ``JES: Relative Balance'' accounting for the difference in modeling of missing transverse momentum.
}
\label{tab:syst_overview}
\renewcommand{\arraystretch}{1.1}
\begin{tabular}{c c c c}
    Uncertainty category & Type & Effect on yield & Effect on shape \\
    \hline
    \multicolumn{4}{c}{\textit{Experimental uncertainties}} \\
    Trigger efficiency & shape & 0.5\% & 0.2\% \\
    Lepton ident./isolation & shape & 3.0\% & 0.2\% \\
    \PQb tagging (heavy) & shape & 0.6\% & 0.1\% \\
    \PQb tagging (light) & shape & 0.3\% & 0.2\% \\
    Kinematic reconstruction & shape & 0.3\% & 0.1\% \\
    JES: Absolute & shape & 0.9\% & 0.3\% \\
    JES: Absolute (stat) & shape & 0.4\% & 0.2\% \\
    JES: Pileup & shape & 0.5\% & 0.2\% \\
    JES: Flavor QCD & shape & 0.7\% & 0.2\% \\
    JES: Relative balance & shape & 0.4\% & 0.8\% \\
    JER & shape & 0.3\% & 0.3\% \\
    Unclustered energy & shape & 0.2\% & 0.3\% \\
    Pileup & shape & 0.4\% & 0.1\% \\
    \ttbar normalization & norm. & 4.4\% & 0.3\% \\
    \Zjets normalization & norm. & 1.6\% & 0.4\% \\
    \Zjets shape & shape & 0.2\% & 0.2\% \\
    Luminosity & norm. & 1.2\% & $<$0.1\% \\[\cmsTabSkip]
    \multicolumn{4}{c}{\textit{Model uncertainties}} \\
    Matrix-elem. renorm. scale variation & shape & 0.4\% & 0.3\% \\
    Matrix-elem. fact. scale variation & shape & 0.6\% & 0.2\% \\
    Parton shower: Initial-state radiation & shape & 0.8\% & 0.7\% \\
    Parton shower: Final-state radiation & shape & 2.1\% & 0.4\% \\
    Top quark mass & shape & 2.4\% & 0.5\% \\
    ME/parton shower matching & norm. & 0.8\% & $<$0.1\% \\
    Underlying event & norm. & 0.8\% & $<$0.1\% \\
    PDF & shape & 0.9\% & 0.1\% \\
    Color reconnection & norm. & 0.8\% & $<$0.1\% \\
    \PQb quark fragmentation & shape & 0.7\% & 0.4\% \\
    \PB hadron semilept. decays & shape & 0.3\% & 0.2\% \\
    Branching fraction & norm. & 1.9\% & $<$0.1\% \\
    NNLO QCD reweighting & shape & 0.6\% & 0.4\% \\
    EWK corrections & shape & 0.6\% & 0.4\% \\
    \PGht normalization & norm. & 0.7\% & 0.8\% \\
    \PGht binding energy & shape & 0.2\% & 0.1\% \\
\end{tabular}
\end{table*}

\subsection{Experimental uncertainties}

{\tolerance=800
In order to improve the description of the data by the simulation, efficiency SFs ($\mathrm{SF}=\epsilon_{\text{DATA}}/\epsilon_{\text{MC}}$) are applied where necessary. Most SFs are applied as event weights to the simulated samples.
\par}

Trigger efficiencies (for the OR of the dilepton and single-lepton triggers) are measured as a function of the lepton $\eta$ using triggers that are only weakly correlated with the dilepton triggers (\ptmiss-based triggers).
Measurements from data are used to improve predictions from \Zjets simulation. The measured SFs are typically close to unity across all bins.
Uncertainties related to trigger efficiency SFs are measured using two kinds of variations, following the same procedure as in Ref.~\cite{CMS:2018adi}.
First, the SFs are varied according to their uncertainties in order to account for the change in total rates with respect to the nominal simulation. The second kind of variation accounts for the shape modification.
This is achieved by varying the SFs with ``up'' and ``down'' variations, based on the different combinations of dilepton $\eta$ regions. An envelope of the two variations is constructed to determine the uncertainty.

{\tolerance=800
Lepton identification and isolation efficiencies (referred to as ``Lepton ident./isolation'') for electrons and muons are estimated as a function of \pt and $\eta$ using a ``tag-and-probe'' method~\cite{CMS:2015yky, CMS:2011hqo} based on the \PZ boson samples.
The SFs are found to be close to unity and they are varied according to their uncertainties.
\par}

Tagging efficiencies for \PQb, \PQc, and light-quark or gluon jets (referred to as ``\PQb tagging (heavy)'' and ``\PQb tagging (light)'', respectively) are determined from simulation as a function of jet \pt and $\eta$~\cite{CMS:2017wtu}. Since the \PQb tagging efficiencies measured in data are different from those measured in simulation, SFs are applied to the simulated events.
Envelopes of the normalization and shape uncertainties are calculated for both the heavy and light flavors and these are used independently as input templates to the profile likelihood fit.

For a fraction of events (about 9\%), the \ttbar kinematic reconstruction finds no physical result as the only solutions to the quadratic equation are imaginary and the events are rejected. This occurs because of resolution effects and the top quark and \PW boson mass constraints.
The efficiency is expressed as a function of lepton observables (\pt and $\eta$), the number of \PQb jets, and \ptmiss.
The data-to-simulation SFs are derived separately for the \ee, \mumu, and \emu final states. Uncertainties are estimated following the procedure in Ref.~\cite{CMS:2018adi}.

The uncertainties due to the jet energy scale (JES) are determined from 19 JES sources~\cite{CMS:2016lmd}.
These sources arise from differences due to variations of the energy of the reconstructed jets (``JES: Absolute'' and ``JES: Absolute (stat)'') and pileup offset dependence (``JES: pileup'') in simulated events between different detector regions, as well as due to corrections for initial- (ISR) and final-state radiation (FSR), and also due to uncertainties in the correction factors of heavy- and light-flavor jets (``JES: Relative samples'' and ``JES: Flavor QCD''). ``JES: Relative Balance'' accounts for the difference in modeling of missing transverse momentum. Simulations are run with jet four-momenta scaled by the corresponding uncertainties to obtain ``up'' and ``down'' systematic samples. Variations of the jet four-momenta are propagated to \ptmiss measurements.
The selection efficiency is recalculated with rescaled simulated samples and the difference with respect to the original samples is taken as systematic uncertainty for the relevant JES source.

The jet energy resolution (JER) is smeared using a scaling method whereby the difference between the reconstructed jet \pt and the matched generated jet \pt is scaled by a certain factor. Uncertainties in the correction factors are evaluated accordingly. The simulated JER is varied by $\pm 1 \sigma$ uncertainty in different $\eta$ regions~\cite{CMS:2016lmd}.

The energy deposited by charged and neutral hadrons and photons is varied according to their energy resolutions, and the \ptmiss four-vectors are recalculated to account for the unclustered \ptmiss (referred to as ``Unclustered energy'').

The pileup effect on the signal efficiency is accounted for by varying the total inelastic cross section by $\pm 4.6\%$ with respect to the nominal value of 69.2\unit{mb}~\cite{CMS:2018mlc}.

{\tolerance=800
The normalization of the background events represents another source of systematic uncertainty, estimated by using the corresponding experimental uncertainties per background source. This uncertainty is assigned a log-normal prior.
In the \ee and \mumu channels, the background from \Zjets processes is dominant. Its normalization (referred to as ``\Zjets normalization'') is varied by $\pm20\%$~\cite{CMS:2015yky}.
A data-driven approach is applied to estimate this uncertainty where the MC prediction in the \PZ boson mass region is compared to data at different points during the selection, which yields a SF of up to 1.2 applied to the \Zjets sample, and the full size of that SF is taken as a 20\% normalization uncertainty. In addition, a shape uncertainty (referred to as ``\Zjets shape'') is determined to cover differences in modeling \Zjets production at LO and NLO. The NLO \Zjets sample is used to determine a slope in the \cosphi distribution and a $\pm5$\% variation is applied to derive up/down templates.
\par}

Finally, the uncertainty in the integrated luminosity for 2016 data (1.2\%)~\cite{CMS-LUM-17-003} is propagated to the normalization of all simulated predictions including \ttbar and \PGht.

\subsection{Model uncertainties}

The impact of theoretical assumptions on the modeling is determined by considering uncertainties that affect both the normalization and shape and those that affect only the normalization. The former are parametrized by nuisances with Gaussian priors and the latter by nuisances with log-normal priors.

Renormalization and factorization scales in the ME calculation in \Powhegvtwo samples (referred to as ``Matrix-elem. renorm. scale variation'' and ``Matrix-elem. fact. scale variation'', respectively) are varied considering the following variations as a function of the \cosphi distribution:
\begin{itemize}
\item $\mu_{\mathrm{R}}$ is fixed, $\mu_{\mathrm{F}}$ is varied by 2.0 (0.5) for the up (down) shape;
\item $\mu_{\mathrm{F}}$ is fixed, $\mu_{\mathrm{R}}$ is varied by 2.0 (0.5) for the up (down) shape;
\end{itemize}
To evaluate the impact of the choice of the \alpS value in the parton shower (PS) simulation, dedicated \PowhegPythia samples are used.

The uncertainties associated with the modeling of ISR and FSR are estimated with the separate up/down variations of the $\alpS^{\mathrm{ISR}}$ or $\alpS^{\mathrm{FSR}}$ parameter in the dedicated samples, changing their scale individually up and down by factors of 2 and $\sqrt{2}$, respectively.
For ISR and FSR uncertainties we use the up and down variations to determine a template representing a shape uncertainty from a linear fit to these variations. The slope parameters determined from these linear fits to the \cosphi distribution are $-0.0078\pm0.0060$ (ISR) and $-0.0021\pm0.0033$ (FSR).
From these slope parameters, we derive symmetrized templates representing a shape uncertainty due to ISR and FSR effects (referred to as ``Parton shower: Initial-state radiation'' and ``Parton shower: Final-state radiation'', respectively).

Two \Powhegvtwo samples are used to estimate the impact of the value used for \mtop in the MC sample on signal selection efficiency. Compared to the \ttbar\ \Powhegvtwo sample, these two have identical settings but different values of \mtop and are used to derive templates corresponding to the down and up variations of \mtop by 0.5\GeV. 

Two \MGaMCatNLOOnly\ \PGht samples are used to estimate the impact of \mtop in the \PGht process. Also, identical settings are used for these \PGht samples except for \mtop. These samples are used to derive templates corresponding to the down and up variations of \mtop by 0.5\GeV. 

To account for the uncertainty due to ME-PS matching (referred to as ``ME/parton shower matching''), the $h_{\mathrm{damp}}$ parameter is varied up and down by one $\sigma$ with respect to its default value of $h_{\mathrm{damp}}=1.58^{+0.66}_{-0.59}\,\mtop$ in the \PowhegPythia simulation~\cite{CMS-PAS-TOP-16-021}.
As no significant shape effect is observed but the linear fit yields a normalization effect of typically $<$1\%, we model this effect as a 1\% uncertainty with a log-normal prior in the \ttbar normalization.

The uncertainty due to the underlying event tune (referred to as ``Underlying event'') is measured using the \PowhegPythia simulations with up/down variations of the tune parameters with respect to its default value according to their uncertainties~\cite{CMS:2019csb}.
Again, a linear fit of the \cosphi distribution of these up/down variations is utilized to determine a shape uncertainty. A $<$1\% flat effect is observed
and hence, is assigned an uncertainty of 1\% with a log-normal prior.

The uncertainty arising from the PDFs is assessed by reweighting the \ttbar signal sample according to the 100 replicas in the NNPDF3.0 PDF set.
Also, the \alpS value used in the NNPDF3.0 PDF set is varied separately according to its uncertainties.

A multiple particle interaction (MPI) scheme with early resonance decays switched off is implemented in the \Pythia simulation of the \ttbar process as a default color reconnection model. Various color reconnection models (MPI-based scheme with early resonance switched on, gluon-move scheme, and QCD-inspired scheme)~\cite{CR-QCD-inspired, CR-model} are applied to the signal sample independently.
Since no significant shape effect is observed, a log-normal prior on the \ttbar normalization of 1\% is assigned as uncertainty for color reconnection.

The Bowler--Lund function~\cite{Bowler, Lund} with a parameter value of 0.855~\cite{CMS:2021sln} is used as a default fragmentation function to describe momentum transfer from \PQb quarks to \PB hadrons in the \ttbar simulation.
The effect of an up and down variation of the Bowler--Lund function is studied by reweighting the relevant transfer function at the generator level of the \ttbar sample.
Furthermore, a different fragmentation function (Peterson function~\cite{ParticleDataGroup}) is used as an alternative function to reweight the signal sample. A shape dependent total uncertainty due to the fragmentation model (referred to as ``\PQb quark fragmentation'') is estimated by taking an envelope of the results from the two approaches.

The semileptonic branching fraction (BF) of \PB hadrons could affect the \PQb jet energy response. The \PQb jets with semileptonic \PB hadron decays (referred to as ``\PB hadron semilept. decays'') are reweighted at the generator level by comparing the \Pythia BFs with the ones taken from PDG 2022~\cite{ParticleDataGroup} and their uncertainties. More details on these effects are provided in Ref.~\cite{Hayrapetyan:2890557}. This effect results in a shape uncertainty.

The combined dileptonic BF for the \ttbar decay into \ee, \mumu, and \emu final states including leptonic decays of \PGt leptons is $\mathcal{B}_{\text{combined}}=0.06425$ with an uncertainty (referred to as ``Branching fraction'') of 1.8\%~\cite{ParticleDataGroup}, and this is assigned a log-normal prior.

We reweight the nominal signal MC sample such that the top quark \pt spectrum matches that from a fixed-order ME calculation at NNLO in QCD. The nominal signal MC sample without this NNLO QCD reweighting is taken as the systematic variation in the NNLO QCD reweighting. Since this is a one-sided variation, we supply this as the positive variation in the fit and restrict the fit to only consider the positive variation.

An uncertainty in the EWK corrections is included to account for contributions from higher mixed-order terms, following the same prescription as outlined in Ref.~\cite{CMS:2020djy} and allowing us to estimate the impact of missing higher-order terms in the ME calculations.
The EWK corrections are applied to the \ttbar signal sample using the multiplicative approach and the uncertainty is taken as the difference between the multiplicative and additive approaches. Since this is a one-sided variation, we supply this as the positive variation in the fit and restrict the fit to only consider the positive variation.

We assign specific systematic uncertainties to compensate for the unknown \PGht signal component with respect to its normalization and shape. The expected migration between bins of the \cosphi distribution is obtained from simulated \ttbar events at reconstruction level and used to smear the generator-level events in the simulated \PGht sample.

In order to determine a model uncertainty for the toponium contribution we rely on a theoretical result in Ref.~\cite{Ju:2020otc}. The normalized \mttbar cross section distribution between 300 and 380\GeV, which is predicted at NLO plus next-to-leading power~\cite{Beneke2022} (NLO$+$NLP) and NNLO$+$NLP levels, is compared with earlier results using CMS data~\cite{CMS:2018adi} and shows reasonable agreement when applying a SF of 1.2. However, the toponium model at LO QCD utilized in this measurement only uses the \PGht color-singlet contribution and does not adequately cover effects of color-octet contributions. Hence, we vary the \PGht cross section by $\pm50$\% of its predicted value to account not just for resummation effects, but also for the effects described earlier. The predictions for the binding energy of \PGht under various potentials has been shown to vary from $-1.7$ to $-2.6$\GeV~\cite{martin_toponium_1988}. Thus, we also vary the binding energy of the \PGht by $\pm 0.5\GeV$, which directly influences the formation and stability of the \ttbar bound state and quantifies the energy required to separate the top quark and antiquark constituents, impacting the shape of the \PGht signal component. The choice of the size of this variation ensures that the model returns physical solutions with a negative binding energy in the variation. Missing color-octet contributions are also affected by variations of the binding energy, but this effect is included in the overall normalization uncertainty.

\section{Modeling of the production threshold}
\label{sec:productionThreshold}

\begin{figure*}[!p]
\centering
\includegraphics[width=0.48\textwidth]{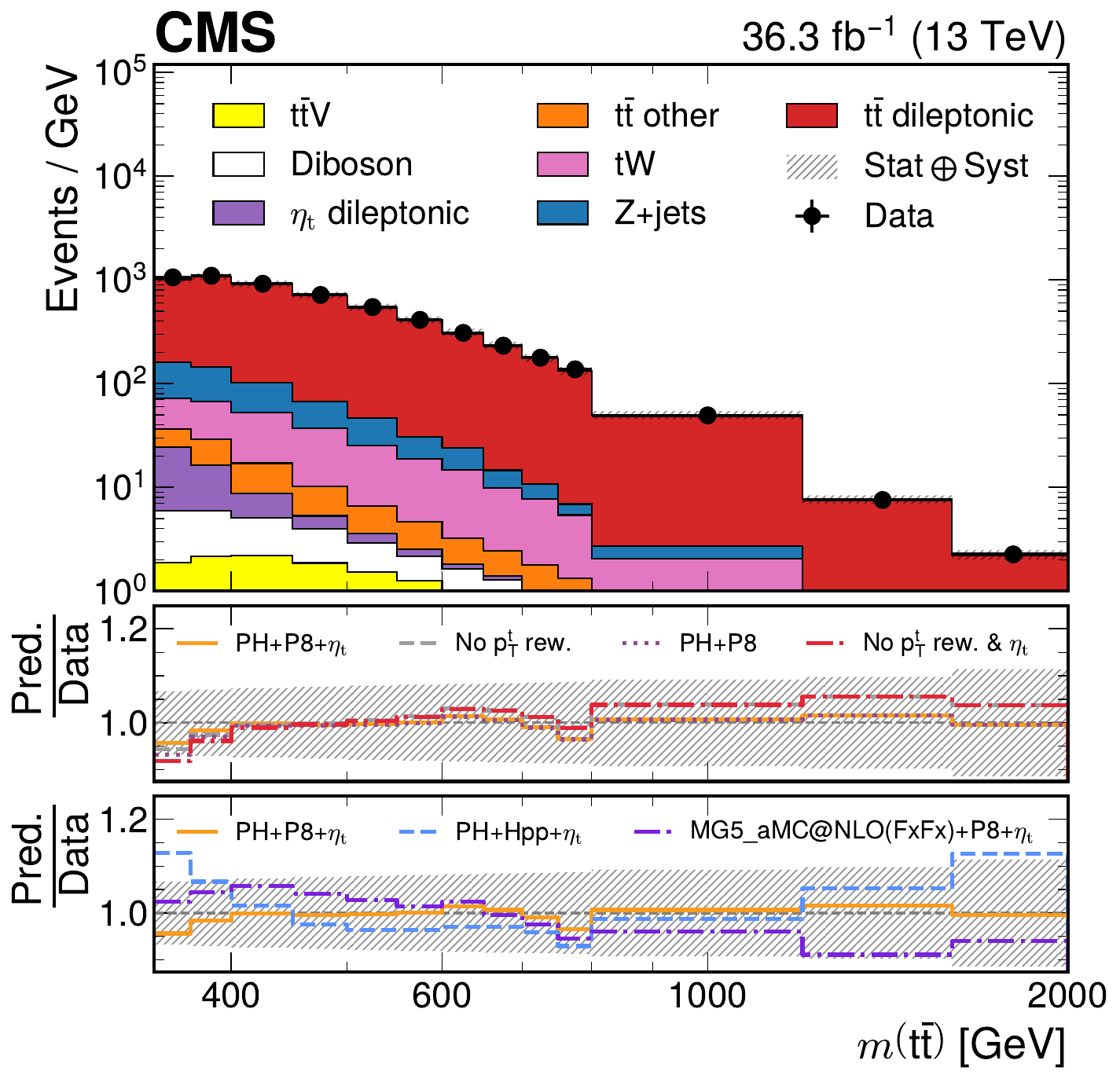}%
\hfill%
\includegraphics[width=0.48\textwidth]{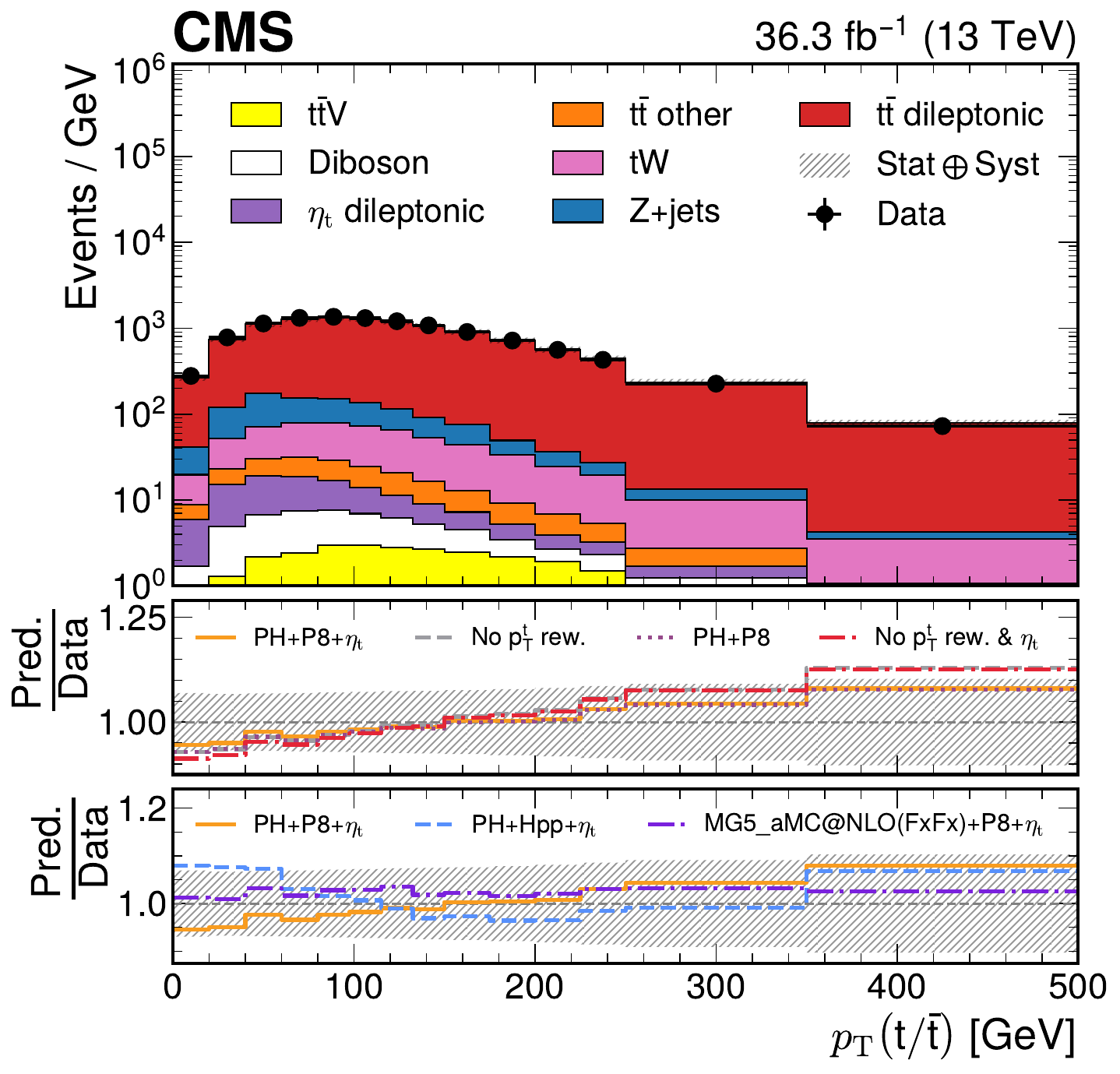} \\
\includegraphics[width=0.48\textwidth]{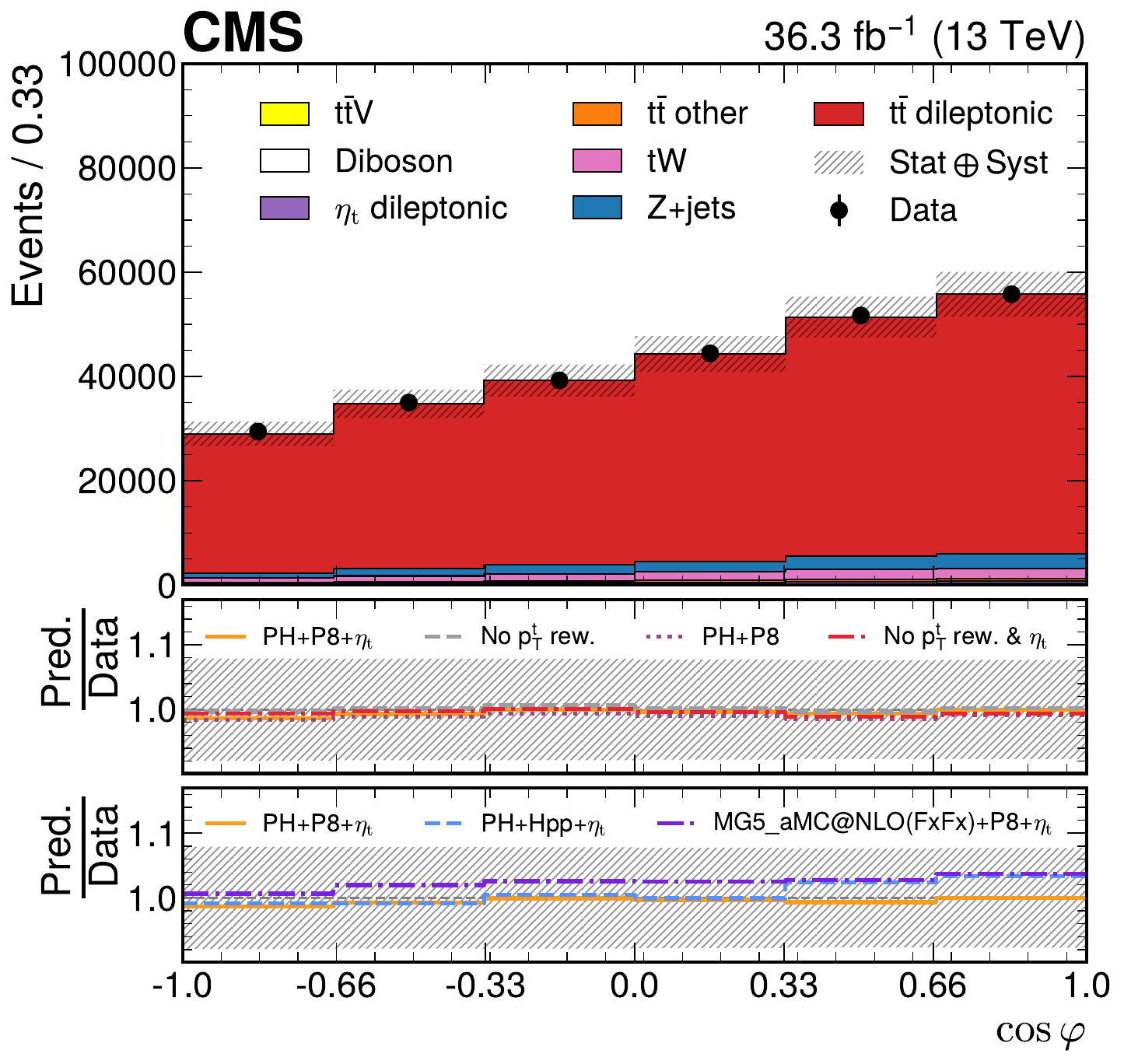}
\caption{%
    Reconstruction-level \mttbar (upper left), \pttort (upper right), and \cosphi (lower) distributions of the combined signal model (\PowhegPythiaEtatLong, labeled \PowhegPythiaEtat) in the full phase space comparing the modeling of the data by MC simulation when not including \PGht contributions (purple dotted line in the upper panel under each plot), or no \pttort reweighting is applied (gray dashed line in the upper panel), or neither of those (red dashed-dotted line in upper panel). The lower panel under each plot compares the data to \PowhegHerwig (blue dashed line, labeled \PowhegHerwigEtat), to \MadgraphPythia (purple dashed-dotted line), and finally to the nominal MC including \PGht contributions and \pttort reweighting (orange solid line, labeled \PowhegPythiaEtat). The hashed uncertainty bands correspond to the pre-fit systematic uncertainties and includes the statistical uncertainty of the data as well. The label ``\pt rew.'' in the legend refers to the \pt reweighting procedure (detailed in Section~\ref{sec:efficiencies}) used to reweight the \ttbar sample to NNLO in QCD.
}
\label{fig:thresholdWtoponium}
\end{figure*}

Several kinematic variables, shown in Fig.~\ref{fig:thresholdWtoponium}, were examined to study the modeling of the data by the MC simulation near the production threshold of \ttbar.
The two lower panels of each plot in Fig.~\ref{fig:thresholdWtoponium} show ratios of the various predictions to the data that more clearly demonstrate the improvement in modeling of the data when including contributions from the \PGht state.
Comparisons of different options on how to construct the combined signal model and generator choices are also included. Namely, data are compared in the upper panel to the case of \PowhegPythia not being reweighted to account for higher order corrections to the \pttort distribution (gray line in upper panel), or not including \PGht (purple line in upper panel), or neither of them (red line in upper panel).
The lower panel compares data to the nominal combined signal provided by \PowhegPythia (orange line in lower panel), to \PowhegHerwig (blue line in lower panel) and to \MadgraphPythia (purple line in lower panel). The hashed band includes all systematic uncertainties described in Section~\ref{sec:efficiencies} for the combined signal model and all background contributions. The best overall modeling
is given by the combined signal model of \TTplusETAT and including \pttort and EWK corrections, while \MadgraphPythia provides an improved modeling in \pttort but worse modeling in \mttbar. However, it should be noted that the systematic uncertainties are at the pre-fit level, which means in particular that the \PGht contribution has systematic uncertainties in excess of 50\%.

\begin{figure*}[!ht]
\centering
\includegraphics[width=0.48\textwidth]{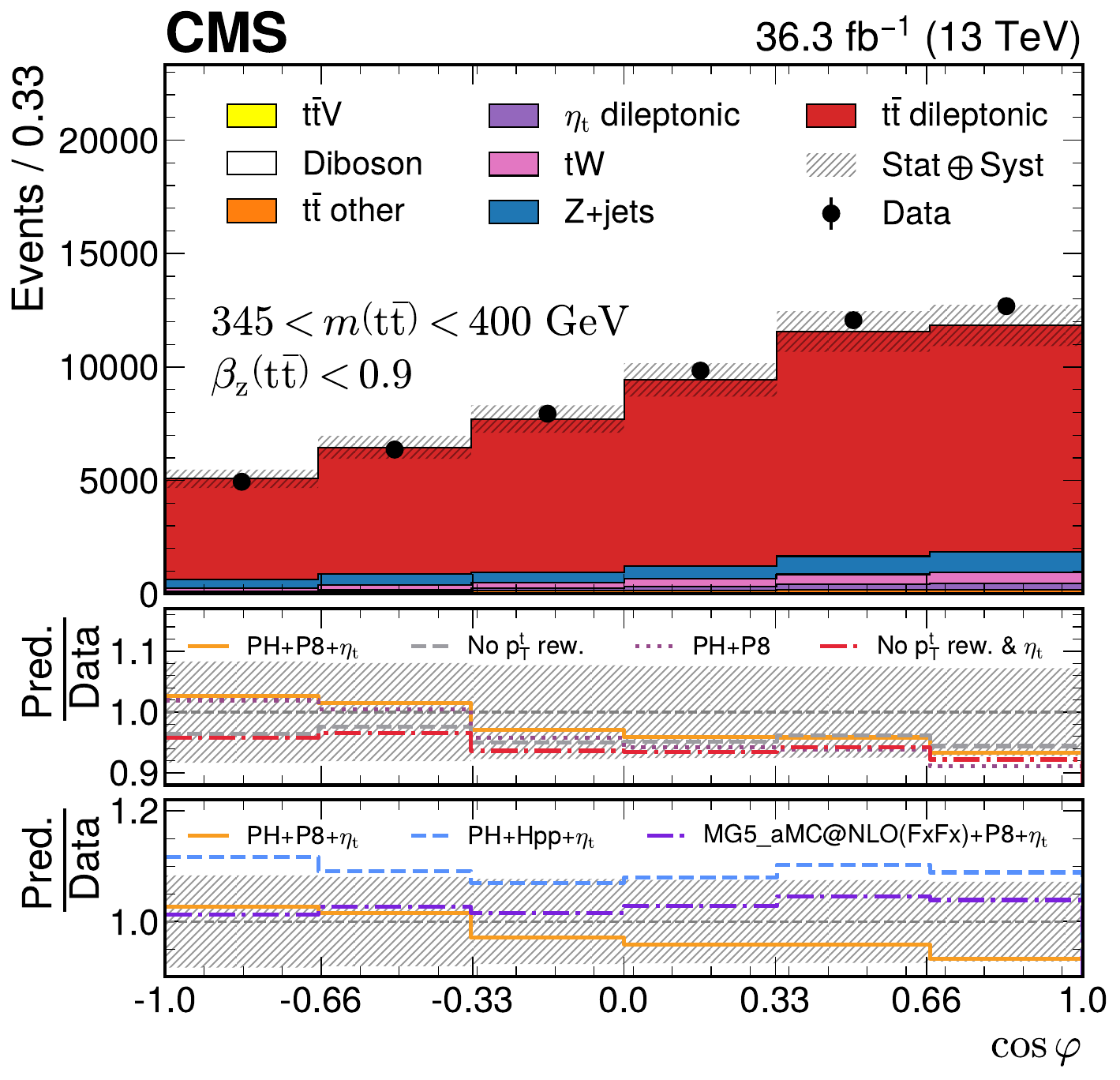}%
\hfill%
\includegraphics[width=0.48\textwidth]{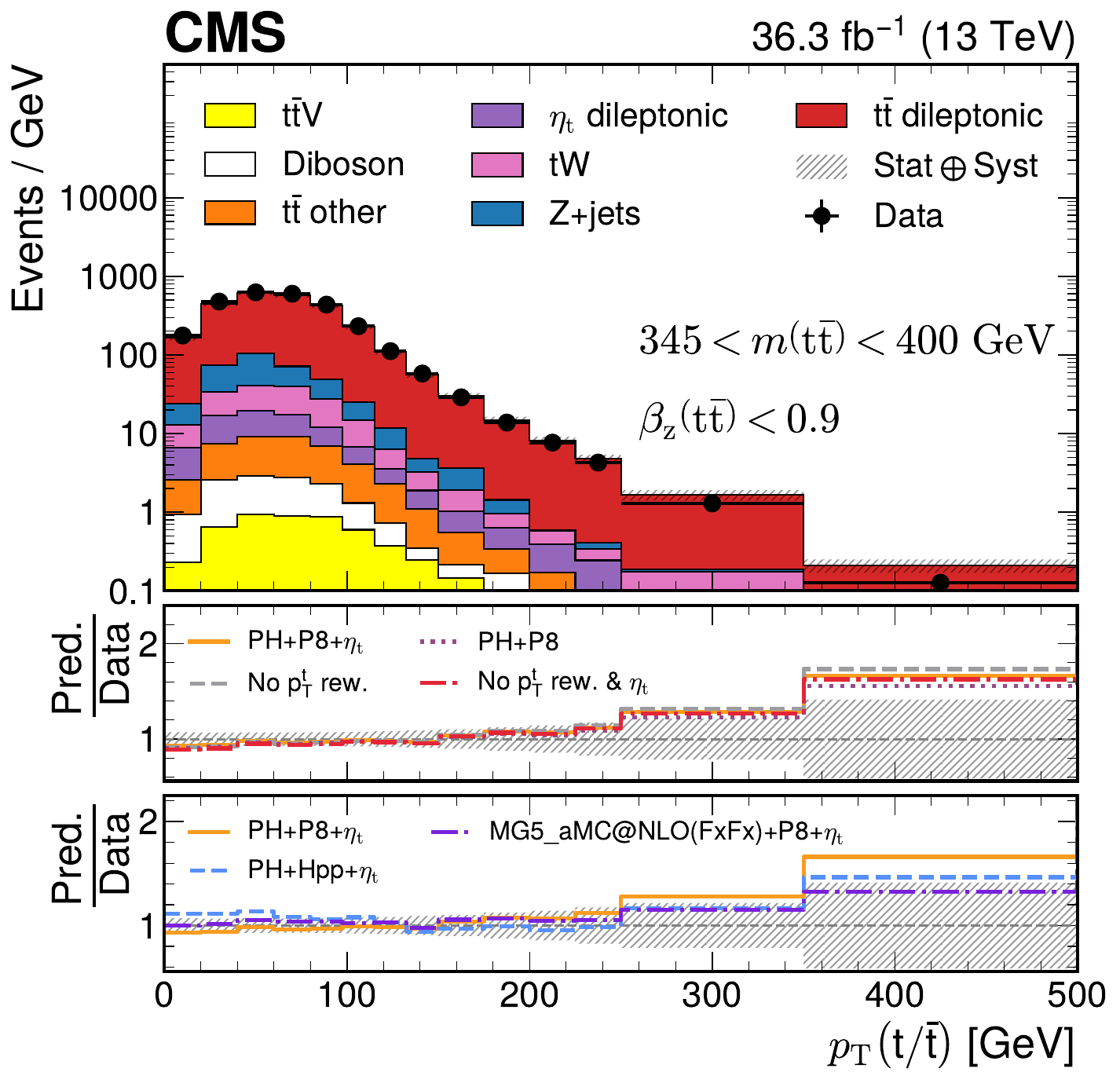}
\caption{%
    Reconstruction-level distributions of \cosphi (left) and \pttort (right) requiring $345<\mttbar<400\GeV$ and $\betaztt<0.9$.
    The lower panels on each figure show the same model comparison done in Fig.~\ref{fig:thresholdWtoponium}. The hashed uncertainty bands correspond to the pre-fit systematic uncertainties and includes the statistical uncertainty of the data as well. The label ``\pt rew.'' in the legend refers to the \pt reweighting procedure (detailed in Section~\ref{sec:efficiencies}) used to reweight the \ttbar sample to NNLO in QCD.
}
\label{fig:recostack}
\end{figure*}

{\tolerance=800
Figure~\ref{fig:recostack} shows MC modeling of the data in the analyzed phase space including the \PGht state residing at lowest \mttbar. Adequate modeling of the data can be seen even in the production threshold region below 400\GeV, which includes the \PGht contribution. Systematic uncertainties were described earlier in more detail in Section~\ref{sec:efficiencies} and include (large) systematic uncertainties for the modeling of the \PGht and the EWK corrections to the signal MC components. Within the selected phase space and including effects from \pttort reweighting and \PGht, \MGaMCatNLO provides a better description than the other event generators for both the \cosphi and the \pttort distributions.
\par}

\section{Extraction of the entanglement proxy}
\label{sec:extractcoefficients}

The distribution of events at the reconstruction-level as a function of \cosphi in the phase space of $345<\mttbar<400\GeV$ and $\betaztt<0.9$ is employed in a binned profile likelihood fit to measure the entanglement proxy $D$ and to compare it with the boundary for entanglement. A single parameter of interest $D$ is defined, which means that the combined \ttbar and \PGht signal are correlated in the likelihood fit, and only a single $D$ is extracted at the end. We use templates with the expected yields as reported in Table~\ref{tab:yields1} as nominal input to the binned profile likelihood fit including shapes of all systematic uncertainties described in Section~\ref{sec:efficiencies} and define $D$ as the single parameter of interest with an unconstrained shape prior. The profile likelihood fit is repeated while varying the entanglement proxy $D$ and the minimum of the negative log likelihood is used to determine the best fit value for $D$. The uncertainty in $D$ is determined from when the difference of the negative log likelihood values to the minimum crosses the values $-2\Delta\ln L=1$.

\begin{figure*}[!ht]
\centering
\includegraphics[width=0.48\textwidth]{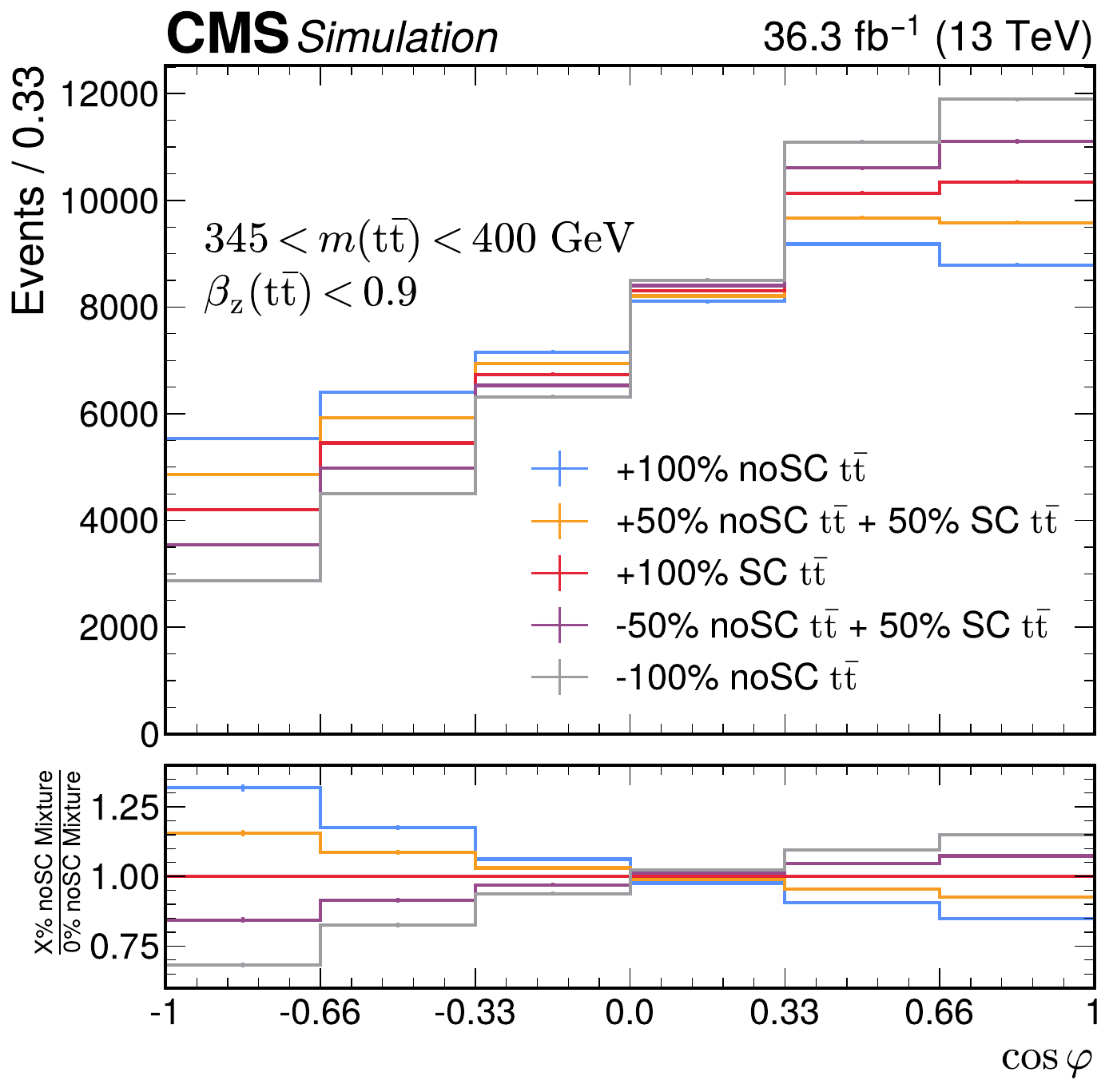}%
\hfill%
\includegraphics[width=0.47\textwidth]{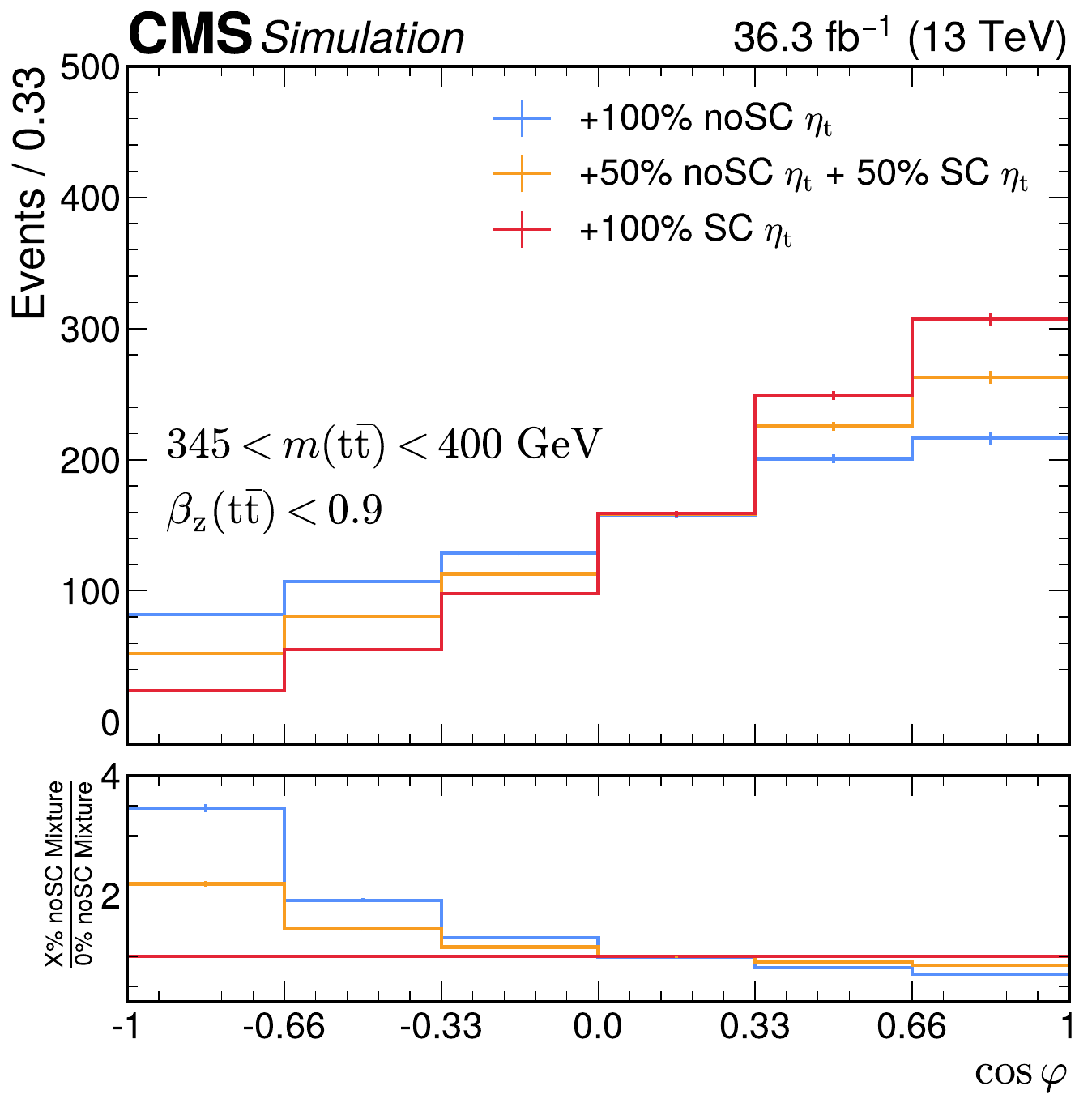}
\caption{%
    Reconstruction-level distribution of the combined \TTplusETAT signal model in mixtures of the noSC combined signal sample. Template variations as a function of \cosphi requiring an \mttbar of $345<\mttbar<400\GeV$ and $\betaztt<0.9$ are shown.
    The \ttbar noSC and SC mixtures ranging from $-100\%$ to $+100\%$ noSC are shown on the left. The \PGht noSC and SC mixtures ranging from zero noSC to $+100\%$ noSC are shown on the right.
}
\label{fig:templates}
\end{figure*}

In particular, templates are derived using the \Powhegvtwo SM predictions implementing spin correlations
for the \ttbar component of the combined signal model including the assumed \PGht contribution. We expect that mixtures of such a sample and a sample with purposely broken spin correlations can effectively model a continuous (linear) variation of the degree of entanglement between the top quarks by means of the entanglement proxy $D$.
In order to have templates implementing an alternative value of the entanglement proxy $D$, we employ the noSC \Powhegvtwo sample (including the assumed \PGht noSC contribution) and ``mix'' it in steps ranging from $-100\%$ to $+100\%$ with the combined signal model SM template. Figure~\ref{fig:templates} shows ``mixtures'' obtained from three values (100\% SC, 50\%/50\% noSC/SC, 100\% noSC) as a function of \cosphi at the detector level for the \ttbar (left) and \PGht (right) components of the combined signal model. The negative mixtures for \ttbar, shown in Fig.~\ref{fig:templates}, are created mirroring the corresponding positive mixtures around the 100\% SC mixture.
The profile likelihood fit is prevented from fitting a negative noSC fraction for the \PGht contribution as this would imply $D < -1$ and would have the consequence of predicting a portion of the differential cross section in \cosphi to be negative, which is non-physical.
Any particular mixture of combined SC and noSC signal corresponds to a certain value of $D$ at the parton level by means of calculating a 2-bin asymmetry ($A_D$) with the combined signal model, consisting of \ttbar and \PGht. The value of
\begin{equation}
    A_D=\frac{N(\cosphi>0)-N(\cosphi<0)}{N(\cosphi>0)+N(\cosphi<0)}
\end{equation}
yields $D$ as $-2A_D$, with $N$ always being the sum of \ttbar and \PGht contributions.

The profile likelihood fit yields the value of the parameter of interest $D$ directly at the parton level accounting for all detector effects due to acceptance, efficiency, and migration. The parton level corrected phase space is given by: $\mttbar<400\GeV$ and $\betaztt<0.9$. These effects smear the definition of when top quark-antiquark states start becoming entangled at the detector level.
We also include the shifts originating from the profile likelihood fit of all systematic uncertainties implemented as nuisance parameters and employ a scan of twice the negative log likelihood ($-2\Delta\ln L$) distribution of the parameter of interest $D$ to measure its value and uncertainty.

The measurement of $D$ via a binned profile likelihood fit and the scan of the $-2\Delta\ln L$ distribution are verified to be unbiased against different values of an injected $D_{\text{inject}}$ at the parton level corresponding to top quarks with a varying degree of entanglement.

In order to determine whether top quarks are inseparable and consequently entangled, the measured value of $D$ can be compared to the boundary for entanglement.

\section{Results}
\label{sec:results}

\begin{figure*}[!ht]
\centering
\includegraphics[width=0.485\textwidth]{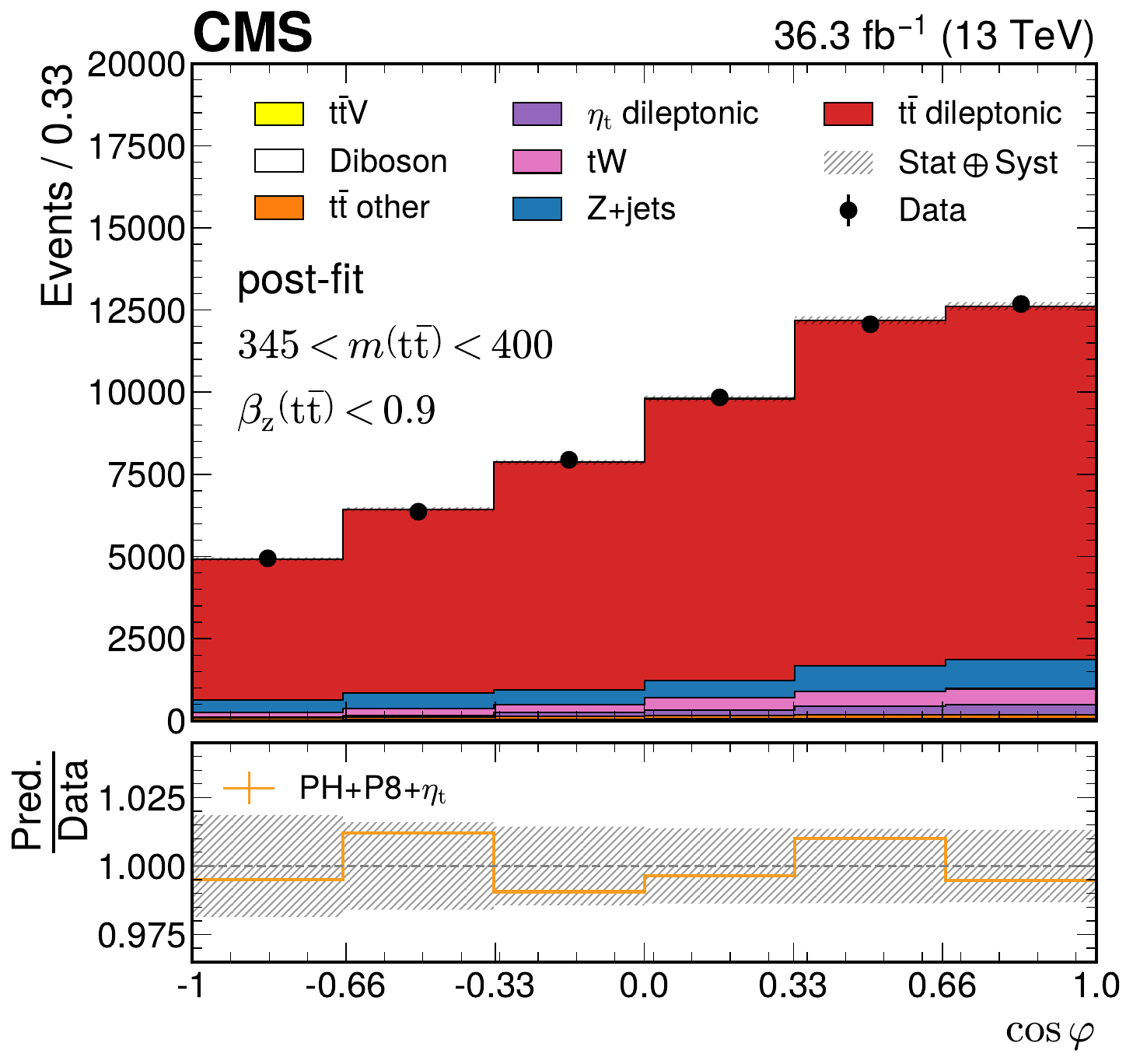}%
\hfill%
\includegraphics[width=0.468\textwidth]{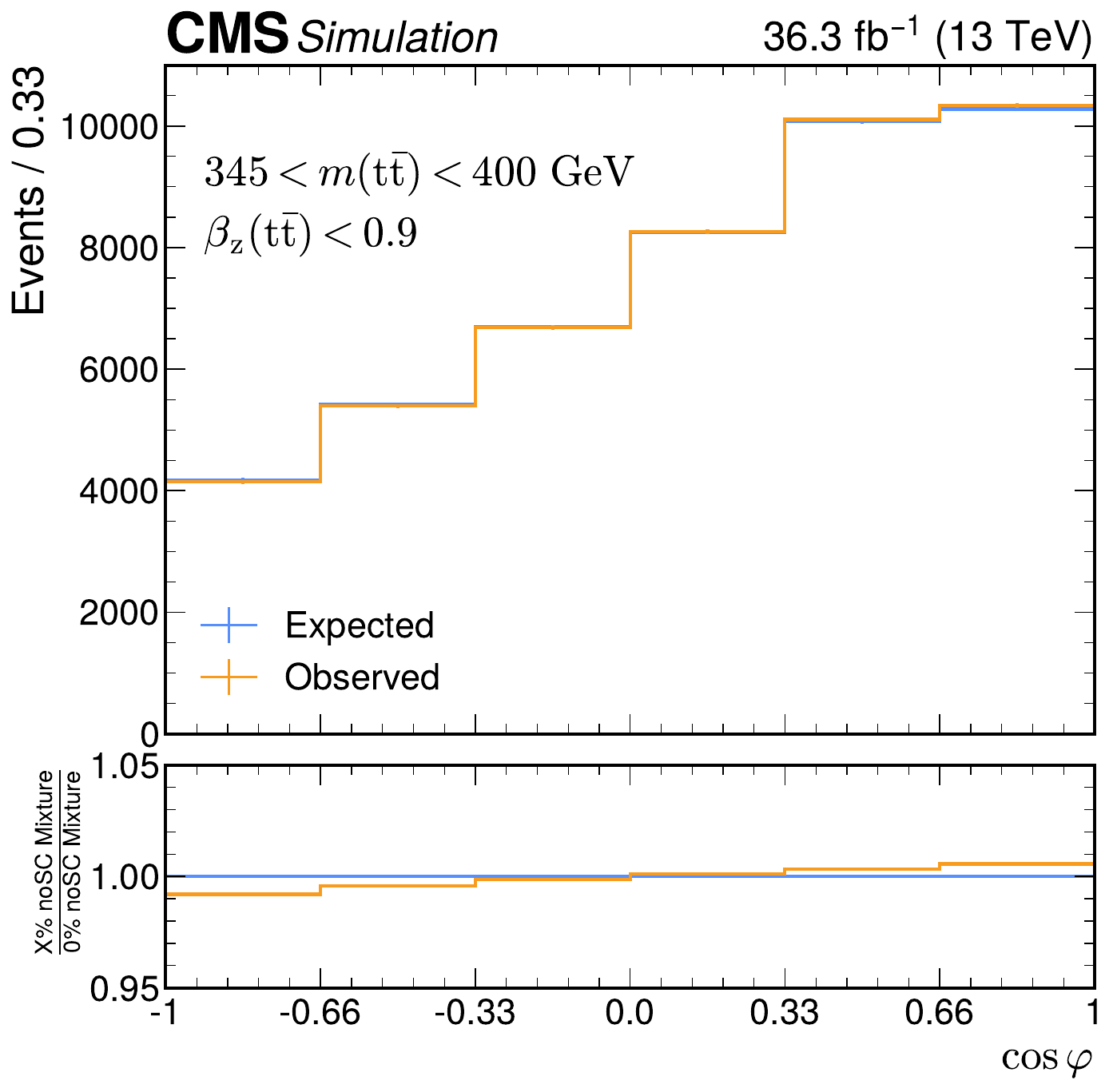}
\caption{%
    The post-fit detector-level distribution of \cosphi requiring $345<\mttbar<400\GeV$, $\betaztt<0.9$, and including \PGht in the fit, is shown on the left. The hashed band corresponds to the post-fit uncertainty and includes the statistical uncertainty of the data added in quadrature. The nominal combined signal model, \PowhegPythiaEtatLong, is labeled as \PowhegPythiaEtat. The fitted noSC and SC mixture template for the combined signal model in the \cosphi distribution is shown on the right.
}
\label{fig:postFit}
\end{figure*}

The result of the binned profile likelihood fit of the \cosphi distribution is shown in Fig.~\ref{fig:postFit} (left), and the data is well modeled by the combined signal model of \TTplusETAT. Figure~\ref{fig:postFit} (right) presents the expected and observed template of noSC and SC mixture and we observe a best fit mixture of the post-fit templates resulting in a \ttbar contribution consistent with a 2.53\% more spin correlated \ttbar contribution when compared to the SM. The \PGht contribution is consistent with 100\% SC contribution, which is the expectation by the SM for the \PGht contribution.

\begin{figure*}[!ht]
\centering
\includegraphics[width=0.48\textwidth]{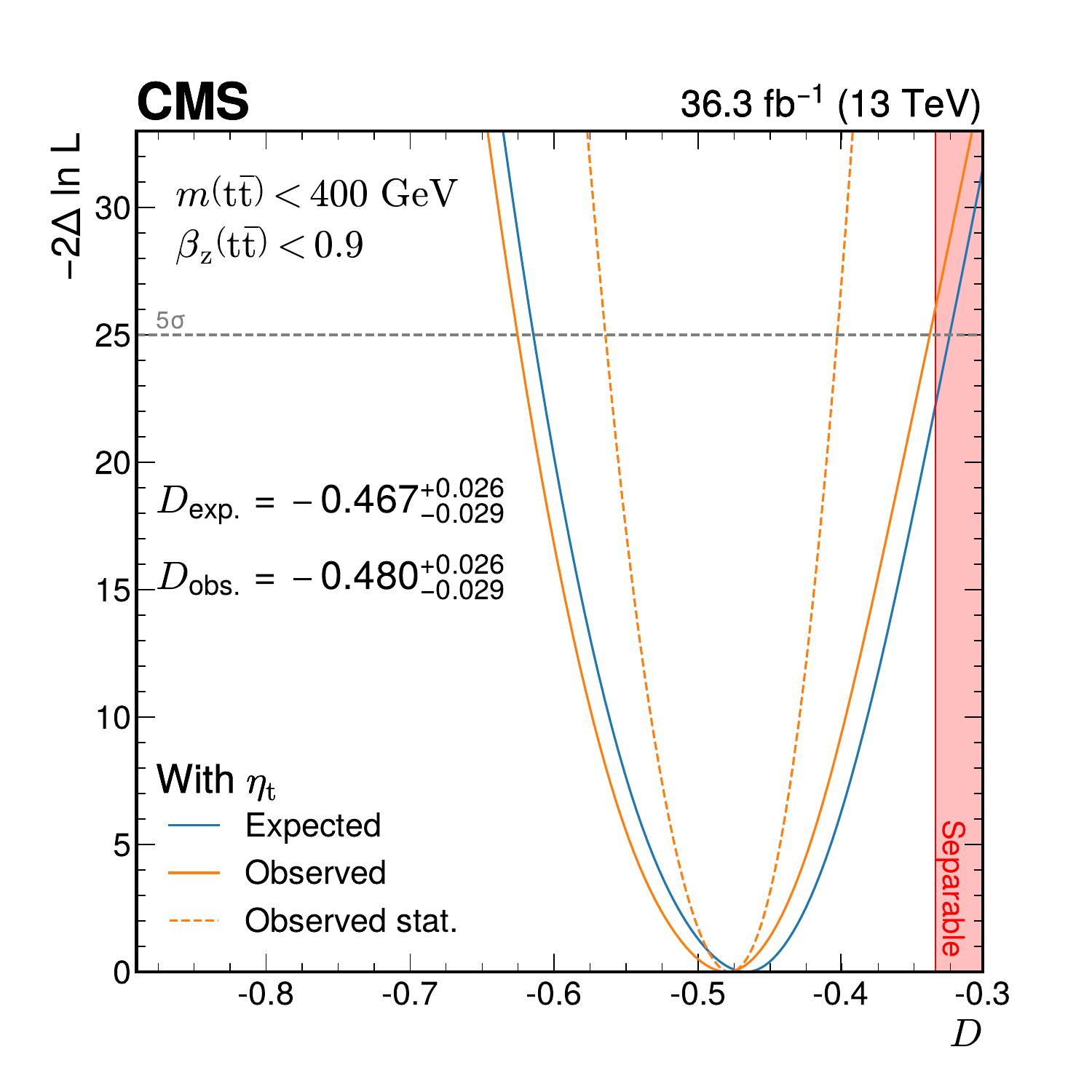}
\includegraphics[width=0.48\textwidth]{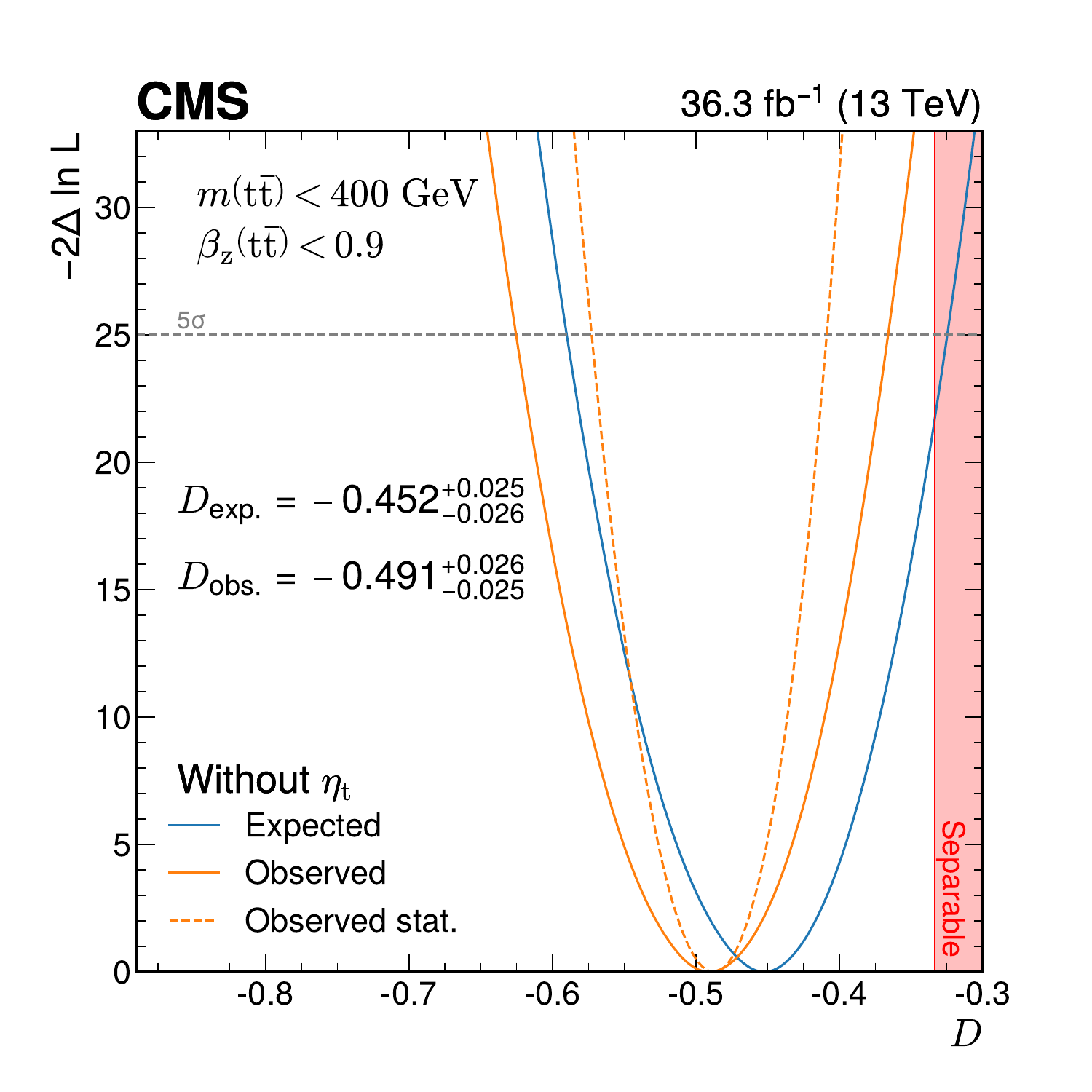}
\caption{%
    Result of the scan of the quantity $-2\Delta\ln L$ from a profile likelihood fit as a function of the parameter of interest $D$, when including (left) or excluding (right) the \PGht contribution. Both results are at parton level and the relevant phase space is indicated in the figures itself. The region where the \ttbar pairs become separable and not entangled ($D>-1/3$) is indicated by the shaded area.
}
\label{fig:lnLScan}
\end{figure*}

Table~\ref{tab:postfit} provides the yields for each simulated sample and data at the pre-fit and post-fit level. The scan of the $-2\Delta\ln L$ distribution of the parameter of interest $D$ is shown in Fig.~\ref{fig:lnLScan} including the boundary for entanglement at $D=-1/3$.

\begin{table}[!ht]
\centering
\topcaption{%
    The number of predicted and observed events in the selected phase space,
    before the fit to the data (pre-fit) and with their best fit normalizations (post-fit). The uncertainties in the pre-fit and post-fit yields reflect total uncertanties but do not include correlations. The ``Only \PGht'' contribution is not added to the total MC prediction since it is included in the combined signal contribution.
}
\label{tab:postfit}
\renewcommand{\arraystretch}{1.2}
\begin{tabular}{c D{,}{\pm}{-1} D{,}{\pm}{-1}}
    Sample
    & \multicolumn{1}{c}{Pre-fit}
    & \multicolumn{1}{c}{Post-fit} \\
    \hline
    Signal (\TTplusETAT) & 45800,1300 & 47560,380 \\
    \Zjets & 3560,280 & 3490,260 \\
    \tW & 1873,72 & 1912,64 \\
    \ttbar other & 572,27 & 590,22 \\
    Diboson & 180,12 & 185,11\\
    \ttV & 100,13 & 102,14\\
    Total MC & 52100,1600 & 53840,220 \\
    Only \PGht & 890,220 & 920,190 \\[\cmsTabSkip]
    Data & $53843$ & $53843$ \\
\end{tabular}
\end{table}

The value of the entanglement proxy $D$ in top quark events at the parton level is measured following the method described in the previous section and is available as a HEPData record~\cite{hepdata}. For the phase space of $\mttbar<400\GeV$ and $\betaztt<0.9$ at the parton level, an observed value of $D=-0.480^{+0.016}_{-0.017}\stat\,^{+0.020}_{-0.023}\syst$ is obtained in data, with an expected value of $D=-0.467^{+0.016}_{-0.017}\stat\,^{+0.021}_{-0.024}\syst$.
With the boundary for entanglement at $-1/3$, this result corresponds to top quarks being entangled in this phase space with an observed (expected) significance of 5.1 (4.7) $\sigma$.

Removing the \PGht contribution from the signal model and only considering the \ttbar component as signal and re-measuring $D$ in the same phase space as before yields an observed (expected) value of $D=-0.491^{+0.026}_{-0.025}\,(\text{tot.})$ ($D=-0.452^{+0.025}_{-0.026}\,(\text{tot.})$) at the parton level with an observed (expected) significance of 6.3 (4.7) $\sigma$. Data are described better when the expected \PGht contribution is included in the signal model.

\begin{figure*}[!ht]
\centering
\includegraphics[width=0.7\textwidth]{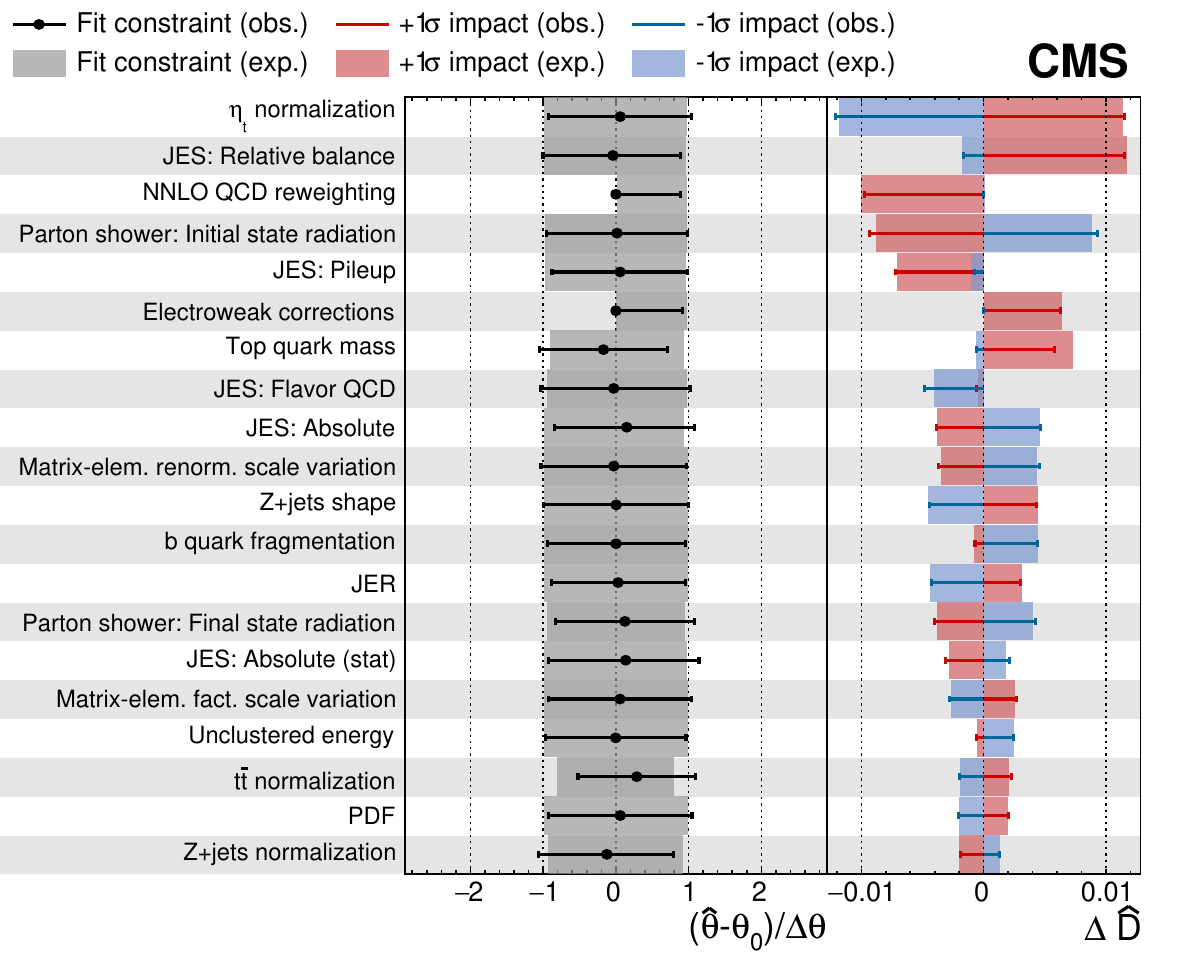}
\caption{%
    The left (right) column shows pulls (impacts) for the top 20 nuisance parameters affecting the measurement of the top quark entanglement. Pulls are calculated using pre- and post-fit values and uncertainties, while impacts are measured by shifting the nuisance parameter by $\pm 1 \sigma$ and observing the change in $D$. The maximum likelihood estimate of the parameter of interest is denoted as $\hat{D}$. The NNLO QCD reweighting and EWK correction uncertainties are one-sided by construction, which can be seen by observing the one-sided pulls shown in the figure. For the NNLO QCD reweighting, we take the uncertainty as the full difference between applying this reweighting and not applying it, and provide it as the positive variation in the profile likelihood fit. For the EWK correction, the uncertainty is taken as the difference between applying the EWK corrections multiplicatively and additively and is provided as the positive variation in the profile likelihood fit.
}
\label{fig:impacts}
\end{figure*}

Figure~\ref{fig:impacts} shows the 20 leading nuisance parameters in the profile likelihood fit. The three leading uncertainties stem from the \PGht signal contribution, the JES relative balance corrections, and the top quark \pt reweighting uncertainty. The latter and the uncertainty on EWK corrections are by construction one-sided.

\begin{figure}[!ht]
\centering
\includegraphics[width=0.48\textwidth]{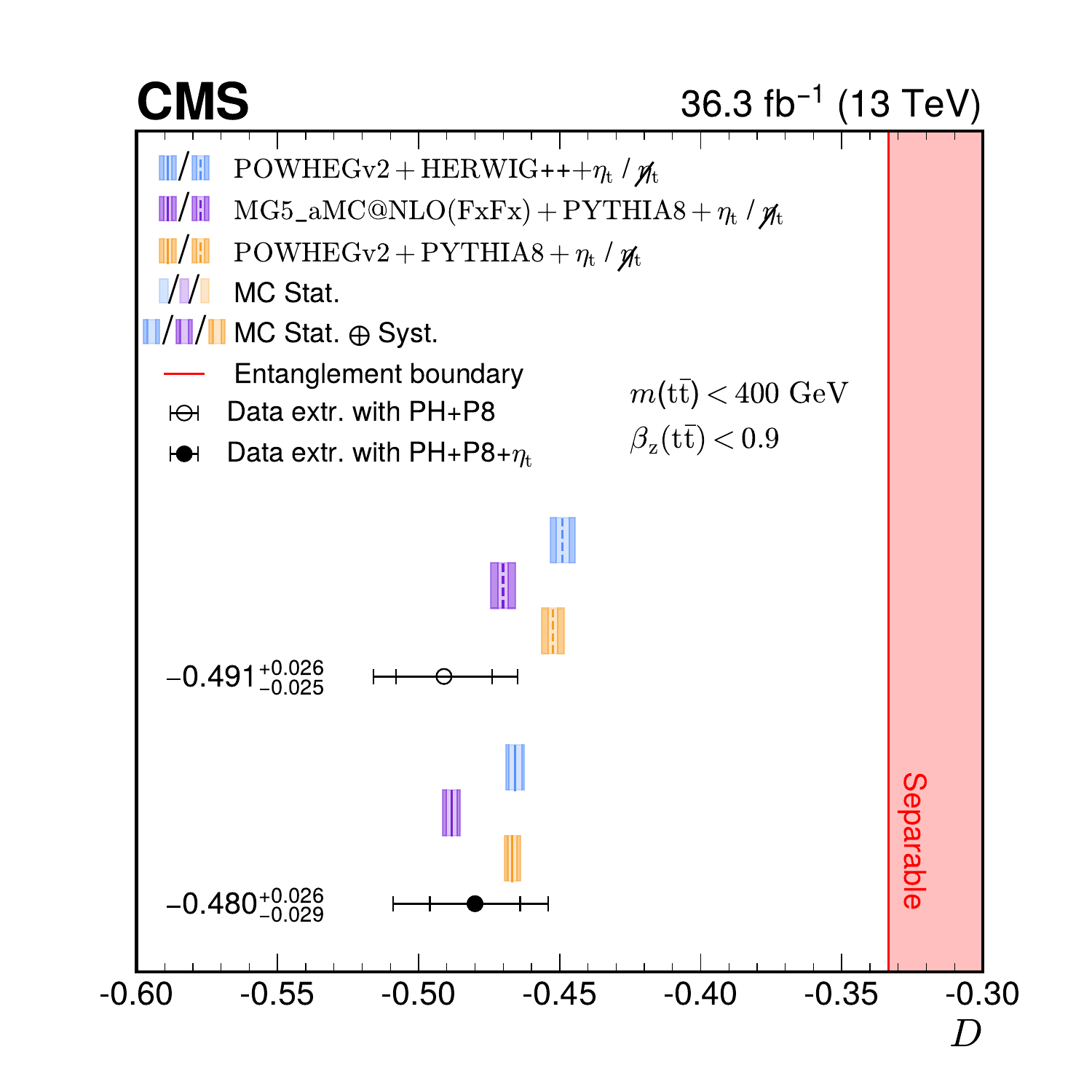}
\caption{%
    Summary of the measurement of the entanglement proxy $D$ in data (black filled or open point) compared with MC predictions including (solid line) or not including (dashed line) contributions from the \PGht state. The legend denotes MC predictions without the \PGht state with a slash through \PGht. Inner error bars represent the statistical uncertainty, while the outer error bars represent the total uncertainty for data. The statistical uncertainty in the MC predictions is denoted by the light shaded region and the total uncertainty, including scale and PDF uncertainties, is represented by the darker shaded region. The boundary for entanglement is indicated by the shaded region at $D=-1/3$.
}
\label{fig:entanglement_results}
\end{figure}

Figure~\ref{fig:entanglement_results} shows the measured value of the entanglement proxy $D$, together with the predicted values from different MC event generators, in the relevant phase space, compared with the boundary for entanglement of top quarks.
Overall, the data is in good agreement with the predictions from the three models, especially with the inclusion of the \PGht contribution.

In addition, Fig.~\ref{fig:entanglement_results} provides the measured and predicted values of $D$ in the same phase space but excluding the \PGht contribution and only assuming the \ttbar signal component. As a result, the measurement shows that \MGaMCatNLO describes the data better, since the predictions by the other two models have some disagreement with the data at the level of one $\sigma$. While the observed values tend to be more negative, the predicted values from all three MCs become less negative owing to the missing effect of the \PGht contribution. This difference in observed value originates from the inclusion of \PGht causing a larger response in the shape of the \cosphi distribution at the reconstruction level for a given change in the parton-level shape when compared with the no spin correlation mixing. Thus, for our specific technique of extracting the entanglement of \ttbar pairs, an overestimation of the observed significance would be obtained, if \PGht contributions were ignored. It is clearly visible that including the \PGht signal component reduces the previously mentioned mild disagreement between data and simulation.

\section{Summary}
\label{sec:summary}

Entanglement is an intrinsic property of quantum mechanics and its measurement utilizes elementary particles to test quantum mechanics. Recently, the ATLAS Collaboration reported the first observation of entanglement in the top quark-antiquark (\ttbar) system~\cite{ATLAS:2023fsd} wih a result indicating a slight deviation from MC simulation.

The measurement of the entanglement of \ttbar pairs performed with CMS data exploits the spin correlation variable $D$, which at the \ttbar production threshold, and in absence of BSM contributions, provides access to the full spin correlation information. This result contrasts with the ATLAS Collaboration's findings in several key ways. We measure entanglement at the parton level, whereas ATLAS reports their observable at the particle level. Additionally, our analysis is the first to consider non-relativistic bound-state effects in the production threshold by including the ground state of toponium, \PGht, which were not included in the ATLAS result. Unlike ATLAS, the CMS result is derived from a binned likelihood fit to extract the entanglement proxy, rather than using a calibration curve.

The $D$ variable represents an entanglement proxy, where a value of less than $-1/3$ signals the presence of entanglement. This proxy is measured using events containing two oppositely charged electrons or muons produced in \pp collisions at a center-of-mass energy of 13\TeV. The modeling of the data is improved when including the additional predicted contribution of the ground state of toponium, \PGht, and is utilized in a combined signal model of \TTplusETAT in the measurement. The extent to which \ttbar pairs are entangled is measured by means of a binned profile likelihood fit of the parameter of interest $D$ directly from the distribution of \cosphi, where $\varphi$ is the angle between the two charged decay leptons in their respective parent top quark rest frames. In the most sensitive kinematic phase space of the relative velocity between the lab and \ttbar reference frames $\betaztt<0.9$, and of the invariant mass of the top quark pair $345<\mttbar<400\GeV$, the fit of the \cosphi distribution yields an observed value of $D=-0.480^{+0.026}_{-0.029}$ and an expected value of $D=-0.467^{+0.026}_{-0.029}$ including the predicted \PGht state.

This result has an observed (expected) significance of 5.1 (4.7) $\sigma$, corresponding to the observation of top quark entanglement. The measured value of $D$ is in good agreement with the MC modeling in this phase space when including the expected \PGht bound state contribution.

\begin{acknowledgments}
We congratulate our colleagues in the CERN accelerator departments for the excellent performance of the LHC and thank the technical and administrative staffs at CERN and at other CMS institutes for their contributions to the success of the CMS effort. In addition, we gratefully acknowledge the computing centers and personnel of the Worldwide LHC Computing Grid and other centers for delivering so effectively the computing infrastructure essential to our analyses. Finally, we acknowledge the enduring support for the construction and operation of the LHC, the CMS detector, and the supporting computing infrastructure provided by the following funding agencies: SC (Armenia), BMBWF and FWF (Austria); FNRS and FWO (Belgium); CNPq, CAPES, FAPERJ, FAPERGS, and FAPESP (Brazil); MES and BNSF (Bulgaria); CERN; CAS, MoST, and NSFC (China); MINCIENCIAS (Colombia); MSES and CSF (Croatia); RIF (Cyprus); SENESCYT (Ecuador); ERC PRG, RVTT3 and MoER TK202 (Estonia); Academy of Finland, MEC, and HIP (Finland); CEA and CNRS/IN2P3 (France); SRNSF (Georgia); BMBF, DFG, and HGF (Germany); GSRI (Greece); NKFIH (Hungary); DAE and DST (India); IPM (Iran); SFI (Ireland); INFN (Italy); MSIP and NRF (Republic of Korea); MES (Latvia); LMTLT (Lithuania); MOE and UM (Malaysia); BUAP, CINVESTAV, CONACYT, LNS, SEP, and UASLP-FAI (Mexico); MOS (Montenegro); MBIE (New Zealand); PAEC (Pakistan); MES and NSC (Poland); FCT (Portugal); MESTD (Serbia); MCIN/AEI and PCTI (Spain); MOSTR (Sri Lanka); Swiss Funding Agencies (Switzerland); MST (Taipei); MHESI and NSTDA (Thailand); TUBITAK and TENMAK (Turkey); NASU (Ukraine); STFC (United Kingdom); DOE and NSF (USA).

\hyphenation{Rachada-pisek} Individuals have received support from the Marie-Curie program and the European Research Council and Horizon 2020 Grant, contract Nos.\ 675440, 724704, 752730, 758316, 765710, 824093, 101115353, 101002207, and COST Action CA16108 (European Union); the Leventis Foundation; the Alfred P.\ Sloan Foundation; the Alexander von Humboldt Foundation; the Science Committee, project no. 22rl-037 (Armenia); the Belgian Federal Science Policy Office; the Fonds pour la Formation \`a la Recherche dans l'Industrie et dans l'Agriculture (FRIA-Belgium); the Agentschap voor Innovatie door Wetenschap en Technologie (IWT-Belgium); the F.R.S.-FNRS and FWO (Belgium) under the ``Excellence of Science -- EOS" -- be.h project n.\ 30820817; the Beijing Municipal Science \& Technology Commission, No. Z191100007219010 and Fundamental Research Funds for the Central Universities (China); the Ministry of Education, Youth and Sports (MEYS) of the Czech Republic; the Shota Rustaveli National Science Foundation, grant FR-22-985 (Georgia); the Deutsche Forschungsgemeinschaft (DFG), under Germany's Excellence Strategy -- EXC 2121 ``Quantum Universe" -- 390833306, and under project number 400140256 - GRK2497; the Hellenic Foundation for Research and Innovation (HFRI), Project Number 2288 (Greece); the Hungarian Academy of Sciences, the New National Excellence Program - \'UNKP, the NKFIH research grants K 131991, K 133046, K 138136, K 143460, K 143477, K 146913, K 146914, K 147048, 2020-2.2.1-ED-2021-00181, and TKP2021-NKTA-64 (Hungary); the Council of Science and Industrial Research, India; ICSC -- National Research Center for High Performance Computing, Big Data and Quantum Computing and FAIR -- Future Artificial Intelligence Research, funded by the NextGenerationEU program (Italy); the Latvian Council of Science; the Ministry of Education and Science, project no. 2022/WK/14, and the National Science Center, contracts Opus 2021/41/B/ST2/01369 and 2021/43/B/ST2/01552 (Poland); the Funda\c{c}\~ao para a Ci\^encia e a Tecnologia, grant CEECIND/01334/2018 (Portugal); the National Priorities Research Program by Qatar National Research Fund; MCIN/AEI/10.13039/501100011033, ERDF ``a way of making Europe", and the Programa Estatal de Fomento de la Investigaci{\'o}n Cient{\'i}fica y T{\'e}cnica de Excelencia Mar\'{\i}a de Maeztu, grant MDM-2017-0765 and Programa Severo Ochoa del Principado de Asturias (Spain); the Chulalongkorn Academic into Its 2nd Century Project Advancement Project, and the National Science, Research and Innovation Fund via the Program Management Unit for Human Resources \& Institutional Development, Research and Innovation, grant B37G660013 (Thailand); the Kavli Foundation; the Nvidia Corporation; the SuperMicro Corporation; the Welch Foundation, contract C-1845; and the Weston Havens Foundation (USA).    
\end{acknowledgments}

\bibliography{auto_generated}
\cleardoublepage \appendix\section{The CMS Collaboration \label{app:collab}}\begin{sloppypar}\hyphenpenalty=5000\widowpenalty=500\clubpenalty=5000
\cmsinstitute{Yerevan Physics Institute, Yerevan, Armenia}
{\tolerance=6000
A.~Hayrapetyan, A.~Tumasyan\cmsAuthorMark{1}\cmsorcid{0009-0000-0684-6742}
\par}
\cmsinstitute{Institut f\"{u}r Hochenergiephysik, Vienna, Austria}
{\tolerance=6000
W.~Adam\cmsorcid{0000-0001-9099-4341}, J.W.~Andrejkovic, T.~Bergauer\cmsorcid{0000-0002-5786-0293}, S.~Chatterjee\cmsorcid{0000-0003-2660-0349}, K.~Damanakis\cmsorcid{0000-0001-5389-2872}, M.~Dragicevic\cmsorcid{0000-0003-1967-6783}, A.~Hoang\cmsAuthorMark{2}\cmsorcid{0000-0002-8424-9334}, P.S.~Hussain\cmsorcid{0000-0002-4825-5278}, M.~Jeitler\cmsAuthorMark{3}\cmsorcid{0000-0002-5141-9560}, N.~Krammer\cmsorcid{0000-0002-0548-0985}, A.~Li\cmsorcid{0000-0002-4547-116X}, D.~Liko\cmsorcid{0000-0002-3380-473X}, I.~Mikulec\cmsorcid{0000-0003-0385-2746}, J.~Schieck\cmsAuthorMark{3}\cmsorcid{0000-0002-1058-8093}, R.~Sch\"{o}fbeck\cmsorcid{0000-0002-2332-8784}, D.~Schwarz\cmsorcid{0000-0002-3821-7331}, M.~Sonawane\cmsorcid{0000-0003-0510-7010}, S.~Templ\cmsorcid{0000-0003-3137-5692}, W.~Waltenberger\cmsorcid{0000-0002-6215-7228}, C.-E.~Wulz\cmsAuthorMark{3}\cmsorcid{0000-0001-9226-5812}
% ...
\par}
\vskip\cmsinstskip
\dag:~Deceased\\
$^{1}$Also at Yerevan State University, Yerevan, Armenia\\
$^{2}$Also at University of Vienna, Vienna, Austria\\
$^{3}$Also at TU Wien, Vienna, Austria\\
% ...
\end{sloppypar}
\end{document}